# Network Medicine Framework for Identifying Drug Repurposing Opportunities for COVID-19

Deisy Morselli Gysi[1,2,*], Ítalo Do Valle[1,*], Marinka Zitnik[3,4,*], Asher Ameli[5,6,*], Xiao Gan[1,2,*], Onur Varol[1,7], Susan Dina Ghiassian[5], JJ Patten[8], Robert Davey[8], Joseph Loscalzo[9], and Albert-László Barabási[1,10]

[1] Network Science Institute and Department of Physics, Northeastern University, Boston, MA 02115, USA;[2] Channing Division of Network Medicine, Department of Medicine, Brigham and Women's Hospital, Harvard Medical School, Boston, MA 02115, USA; [3] Department of Biomedical Informatics, Harvard University, Boston, MA 02115, USA; [4]Harvard Data Science Initiative, Harvard University, Cambridge, MA 02138, USA.[5] Scipher Medicine, 221 Crescent St, Suite 103A, Waltham, MA 02453; [6] Department of Physics, Northeastern University, Boston, MA 02115, USA; [7] Faculty of Engineering and Natural Sciences, Sabanci University, Istanbul 34956, Turkey; [8] Department of microbiology, NEIDL, Boston University, Boston, MA, USA; [9] Department of Medicine, Brigham and Women's Hospital, Harvard Medical School, Boston, MA 02115, USA; [10] Department of Network and Data Science, Central European University, Budapest 1051, Hungary.
[*]Those authors contributed equally

## Abstract

The current pandemic has highlighted the need for methodologies that can quickly and reliably prioritize clinically approved compounds for their potential effectiveness for SARS-CoV-2 infections. In the past decade, network medicine has developed and validated multiple predictive algorithms for drug repurposing, exploiting the sub-cellular network-based relationship between a drug's targets and disease genes. Here, we deployed algorithms relying on artificial intelligence, network diffusion, and network proximity, tasking each of them to rank 6,340 drugs for their expected efficacy against SARS-CoV-2. To test the predictions, we used as ground truth 918 drugs that had been experimentally screened in VeroE6 cells, and the list of drugs under clinical trial, that capture the medical community's assessment of drugs with potential COVID-19 efficacy. We find that while most algorithms offer predictive power for these ground truth data, no single method offers consistently reliable outcomes across all datasets and metrics. This prompted us to develop a multimodal approach that fuses the predictions of all algorithms, showing that a consensus among the different predictive methods consistently exceeds the performance of the best individual pipelines. We find that 76 of the 77 drugs that successfully reduced viral infection do not bind the proteins targeted by SARS-CoV-2, indicating that these drugs rely on network-based actions that cannot be identified using docking-based strategies. These advances offer a methodological pathway to identify repurposable drugs for future pathogens and neglected diseases underserved by the costs and extended timeline of *de novo* drug development.

## Introduction

The disruptive nature of the COVID-19 pandemic has unveiled the need for the rapid development, testing, and deployment of new drugs and cures. Given the compressed timescales, the *de novo* drug development process, that lasts a decade or longer, is not feasible. A time-efficient strategy must rely on drug repurposing (or repositioning), helping identify among the compounds approved for clinical use the few that may also have a therapeutic effect in patients with COVID-19. Yet, the lack of reliable repurposing methodologies has resulted in a winner-takes-all pattern, where more than one-third of registered clinical trials focus on hydroxychloroquine or chloroquine, siphoning away resources from testing a wider range of potentially effective drug candidates.

Drug repurposing algorithms rank drugs based on one or multiple streams of information, such as molecular profiles[1], chemical structures[2], adverse profiles[3], molecular docking[4], electronic health records[5], pathway analysis[6], genome wide association studies (GWAS)[6], and network perturbations[6–14]. As typically only a small subset of the top candidates is validated experimentally, the true predictive power of the existing repurposing algorithms remains unknown. To quantify and compare their predictive power, different algorithms must make predictions for the same set of candidates, and the experimental validation must focus not only on the top candidates, but also on a wider list of drugs chosen independently of their predicted rank.

The COVID-19 pandemic presents both the societal imperative and the rationale to test drugs at a previously unseen scale. Hence, it offers a unique opportunity to quantify and improve the efficacy of the available predictive algorithms, while also identifying potential treatments for COVID-19. Here, we implement three recently developed network-medicine drug-repurposing algorithms that rely on artificial intelligence[14,15], network diffusion[16], and network proximity[10] (Figure 1A, B). To test the validity of the predictions, we identified 918 drugs ranked by all predictive pipelines, that had been experimentally screened to inhibit viral infection and replication in cultured cells[17]. We also collected clinical trial data to capture the medical community's collective assessment of promising drug candidates. We find that the predictive power varies for the different datasets and metrics, indicating that in the absence of *a priori* ground truth, it is impossible to decide which algorithm to trust. We propose, however, a multimodal ensemble forecasting approach that significantly improves the accuracy and the reliability of the predictions by seeking consensus among the predictive methods[14,18].

## Network-based Drug Repurposing

Repurposing strategies often prioritize drugs approved for (other) diseases whose molecular manifestations are similar to those caused by the pathogen or disease of interest[19]. To search for diseases whose molecular mechanisms overlap with the COVID-19 disease, we first mapped the experimentally identified[20] 332 host protein targets of the SARS-CoV-2 proteins (Table S4) to the human interactome[21–24] (Table S3), a collection of 332,749 pairwise binding interactions between 18,508 human proteins (SI Section 1.1). We find that 208 of the 332 viral targets form a large connected component (COVID-19 disease module hereafter, Figure 2B), indicating that the SARS-CoV-2 targets aggregate in the same network vicinity[12,25]. Next, we evaluated the network-based overlap between proteins associated with 299 diseases[26] $(d)$ and the host protein targets of SARS-CoV-2 $(v)$ using the $S_{vd}$ metric[26], finding $S_{vd} > 0$ for all diseases, implying that the COVID-19 disease module does not directly overlap with the disease proteins associated with any single disease (Figure S1-2 and Table S7). In other words, a potential COVID-19 treatment cannot be derived from the arsenal of therapies approved for a specific disease, arguing for a network-based strategy that can identify repurposable drugs without regard for their established disease indication.

We implemented three competing network repurposing methodologies (Figure 1B and SI Section 1.4): i) The artificial intelligence-based algorithm[14,15] maps drug protein targets and disease-associated proteins to points in a low-dimensional vector space, resulting in four predictive pipelines A1-A4, that rely on different drug-disease embeddings. ii) The diffusion algorithm[16] is inspired by diffusion state distance, and ranks drugs based on capturing network similarity of a drug's protein targets to the SARS-CoV-2 host protein targets. Powered by distinct statistical measures, the algorithm offers five ranking pipelines (D1-D5). iii) The proximity algorithm[10] ranks drugs based on the distance between the host protein targets of SARS-CoV2 and the closest protein targets of drugs, resulting in three predictive pipelines of which P1 relies on all drug targets; P2 ignores targets identified as drug carriers, transporters, and drug-metabolizing enzymes; and P3 relies on differentially expressed genes identified by exposing each drug to cultured cells[27]. The low correlations across the three algorithms indicate that the methods extract complementary information from the network (Figure 2C, and SI Section 1.5).

**Experimental and Clinical Validation of Drug Repurposing Pipelines**

We implemented the 12 pipelines to predict the expected efficacy of 6,340 drugs in Drugbank[27] against SARS-CoV-2 and extracted and froze the predictions in the form of 12 ranked lists on April 15, 2020. All pipelines rely on the same input data and to maintain the prospective nature of the study, all subsequent analysis relies on this initial prediction list. As the different pipelines make successful predictions of a different subset of drugs, we identified 918 drugs for which all pipelines (except for P3, which predicts the smallest number of drugs) offer predictions and whose compounds were available in the Broad Institute drug repurposing library[28] (Figure 1), and used two independent datasets to quantify the predictive power of each pipeline over the same set of drugs:

(1) As the first ground truth we used 918 compounds that had been experimentally screened for their efficacy against SARS-CoV-2 in VeroE6 cells, kidney epithelial cells derived from African green monkey[17] (see SI Section 2). Briefly, the VeroE6 cells were pre-incubated with the drugs (from 8 μM down to 8 nM) and then challenged with wild type SARS-CoV-2 strain USA-WA1/2020. Of the 918 drugs, 806 had no detectable effect on viral infectivity (N drugs, 87.8% of the tested list); 35 were cytotoxic to the host cells (C drugs); 37 had a strong effect (S drugs), being active over a broad range of concentrations; and 40 had a weak effect (W drugs) on the virus (Figure 3A, Table S10). As the prediction pipelines offer no guidance on the magnitude of the in vivo effect, we consider as positive outcomes drugs that had a strong or a weak effect on the virus (S&W, 77 drugs, Table 2), and as negative outcomes the drugs without detectable effect (N, 806 drugs).

(2) On April 15, 2020 (prediction date), we scanned clinicaltrials.gov, identifying 67 drugs in 134 clinical trials for COVID-19 (CT415 dataset, Table S12). To compare outcomes across datasets, we limit our analysis to the experimentally tested 918 drugs, considering as positive the 37 drugs in clinical trial on the E918 list, and negative the remaining 881 drugs. As the outcomes of these trials are largely unknown, validation against CT415 dataset tests each pipeline's ability to predict the pharmacological consensus of the medical community on drugs with expected potential efficacy for COVID-19 patients.

For the E918 experimental outcomes (Figure 4A), the best area under the curve (AUC) of 0.63 is provided by P1, followed by D2 (AUC = 0.58) and P3 (AUC = 0.58). For CT415 (Figure 4B), we observe particularly strong predictive power for the four AI-based pipelines (AUC of 0.73-0.76), followed by proximity P1 (AUC = 0.57) and P2 (AUC = 0.56).

The goal of drug repurposing is to prioritize all available drugs, allowing the experimental efforts to limit their resources on the top-ranked ones. Thus, the most appropriate performance metric is the number of positive outcomes among top $K$ ranked drugs (precision at $K$), and the fraction of all positive outcomes among the top $K$ ranked drugs (recall at $K$). For the E918 dataset(Figure 4C) A2 ranks 9 S&W drugs among the top 100, followed by P1 (7 drugs) and A3 and A4 (6 drugs). We observe similar trends for recall (Figure 4E): the A2 pipeline ranks 11.7% of all positive drugs in the top 100, and P1 selects 9%. Finally, A1 ranks 12 drugs currently in clinical trials among the top 100 in CT415, followed by A3 (11 drugs) and A2 (10 drugs), trends that are similar for recall (Figure 4F).

Taken together, we find that while most algorithms show statistically significant predictive power (SI Section 3.1, Tables S1-2), they have different performance on the different ground truth datasets: the AI pipeline offers strong predictive power for the drugs selected for clinical trials, while proximity offers better predictive power for the E918 experimental outcomes. While together the twelve pipelines identify 22 positive drugs among the top 100, none of the pipelines offers consistent superior performance for all outcomes, prompting us to develop a multimodal approach that can extract the joint predictive power of all pipelines.

**Multimodal Approach for Drug Repurposing**

Predictive models for drug repurposing are driven by finite experimental resources that limit downstream experiments to those involving a finite number ($K$) of drugs. How do we identify these $K$ drugs to maximize the positive outcomes of the tested list[18]? With no initial knowledge as to which of the $N_p = 12$ predictive pipelines offer the best predictive power, we could place equal trust in all, by selecting the top $K/N_p$ drugs from each pipeline (Union list). We compare this scenario with an alternative strategy that combines the predictions of the different pipelines. A widely used approach is to calculate the average rank of each drug over the $N_p$ pipelines[29] (Average Rank list). The alternative is to search for consensus ranking that maximizes the number of pairwise agreements between all pipelines[15,18]. As the optimal outcome, called the Kemeny consensus[29], is NP-hard to compute, we implemented three heuristic rank aggregation algorithms (RAAs) that approximate the Kemeny consensus: Borda's count[30], the Dowdall method[31], and CRank[15]. For example, if the resources allow us to test $K = 120$ drugs, we ask which ranked list offers the best precision and recall at 120: the Union list collecting the top 10 predictions from the 12 pipelines; or the top 120 predictions of Average Rank, Borda, Dowdall, or CRank; or the top 120 drugs ranked by an individual pipeline.

We find that Average Rank offers the worst performance, trailing the predictive power of most individual pipelines (Figure 4G-H). The Union List and Dowdall offer better outcomes, but trail behind the best performing individual pipelines (E918, CT415). Borda has a strong predictive performance for E918, but not for CT415. In contrast, CRank, that relies on Bayesian factors, offers a consistently high predictive performance for all datasets and most $K$ values. CRank performs equally well for two other datasets: a manually curated prospective list E74 (described below) and the list of clinical trials updated on 06/15/20 (C615, Figure S8). In other words, we find that CRank extracts the cumulative predictive power of all methods, matching or exceeding the predictive power of the individual pipelines across all datasets. Its persistent performance indicates that an unsupervised multimodal approach can significantly improve the hit rate over individual prediction algorithms. It also suggests that in the absence of a ground truth, the Kemeny consensus, which seeks a ranking with the smallest number of pairwise disagreements between the individual pipelines, represents an effective and theoretically principled strategy when each pipeline carries some predictive power.

**Network Effects**

Most computationally informed drug repurposing methods rely on chemical binding energy minimization and docking patterns, limited to compounds that bind either to viral proteins or to the host targets of the viral proteins[20] (Figure 1C). A good example is remdesivir, a direct-acting antiviral that inhibits viral RNA polymerase[32,33]. In contrast, our pipelines identify drugs that target host proteins to induce network-based perturbations that alter the virus's ability to enter the cell or to replicate within it. An example of such host-targeting drug[34] is dexamethasone, which reduces mortality in COVID-19 patients by modulating the host immune system[35]. We find that only one of the 77 S&W drugs are known to directly target a viral protein binding target: amitriptyline, which targets SIGMAR1, the target of the NSP6 SARS-CoV-2 protein. In other words, 76 of the 77 drugs that show efficacy in our experimental screen are "network drugs", achieving their effect indirectly, by perturbing the host subcellular network. As these drugs do not target viral proteins or their host targets target, they cannot be identified using traditional binding-based methods yet are successfully prioritized by network-based methods.

Searching for common mechanistic or structural patterns that could account for the efficacy of the 77 S&W drugs, we explored their target and pathway enrichment profiles (Figures S6-7), as well as their reported mechanisms of action, failing to identify statistically significant features shared by most S&W drugs. This failure is partly explained by the diversity of the S&W drugs (Table S10), containing antipsychotics (9S & 4W), serotonin receptor agonists (3W), non-steroidal anti-inflammatory drugs (2W), angiotensin receptor blockers (2W), tyrosine kinase inhibitors (5S), statins (1W & 2S), and others.

As CRank extracts its predictive power from the network, we hypothesized that network-based patterns may help distinguish the S&W drugs from the N drugs. Indeed, we find that the targets of the 37 S drugs form a statistically significant large connected component (Z-Score=2.05), indicating that these targets agglomerate in the same network neighborhood. We observe the same pattern for the targets of the 40 W drugs (Z-Score=3.42). The negative network separation between the S and W drug targets ($s_{SW} = -0.69$) indicates that, in fact, the S and the W drugs target the *same* network neighborhood. To characterize this neighborhood, we measured the network-based proximity of the targets of the S, W, and N drug classes to the SARS-CoV-2 targets. We find that compared to random expectation, the N drug targets are far from the COVID-19 module (Figure 3C), while the S and W drug targets are closer to the COVID-19 disease module than expected by chance. The magnitude of the effect is also revealing: the S drug targets are closer than the W drug targets, suggesting that network proximity is a positive predictor of a drug's efficacy.

Taken together, our analyses suggest that S&W drugs are diverse, and lack pathway-based or mechanistic signatures that distinguish them. We do find, however, that S&W drug target the same interactome neighborhood, located in the network vicinity of the COVID-19 disease module, potentially explaining their ability to influence viral activity, and the effectiveness of network-based methodologies to identify them.

## Discussion

A recent in vitro screen[36] of 12,000 compounds in VeroE6 cells identified 100 compounds that inhibit viral infectivity. Yet, only 2 of the 12,000 compounds tested are FDA approved, the rest being in the preclinical or experimental phase, years from reaching patients. In contrast, 96% of the 918 drugs prioritized and screened here are FDA approved, hence should they also show efficacy in human cell lines, could be moved immediately to rapid clinical trials. Brute force screening does, however, offer an important benchmark: Its low hit rate of 0.8% highlights the value in prioritizing resources towards the most promising compounds. Indeed, the unsupervised CRank offers an order of magnitude higher (9%) hit rate among the top 100 drugs, and the top 800 of the 6,340 drugs prioritized by CRank contains 58 of the 77 S&W drugs (Figure 4G-H). The hit rate can be further increased by expert knowledge. To demonstrate this point, we mimicked the traditional drug repurposing process whereby a physician-scientist manually inspected the top 10% of the CRank consensus ranking on April 15, removing drugs with known significant toxicities in vivo and lower-ranked members of the same drug class, and arrived at 74 drugs available for testing. Using the experimental design as described above, but over a wider range of doses (0.625 – 20µM, 0.2 MOI), we screened these 74 compounds separately from the E918 list, and found 39 N, 10 W, and 11 S outcomes (Table S11). The resulting 28% enrichment of S&W drugs suggests that in the case of limited resources, outcomes are maximized by combining algorithmic consensus ranking with expert knowledge.

Inspecting the CRank list and the experimental outcomes, we find two highly ranked drugs with strong outcomes, but not yet in clinical trials (Table 1): azelastine (CRank #10, S), an antihistamine used to treat upper airway symptoms of allergies, and digoxin (CRank #33, S), used to treat heart failure. Our findings, coupled with extensive experience in their use in the clinical community, argue for their consideration in clinical trials. Other highly ranked candidates include folic acid (CRank #16, S), or methotrexate (CRank #32, S), which impairs folate metabolism and attenuates host inflammatory response in autoimmune diseases. This latter mechanism argues that methotrexate is likely to be effective at the other end of the disease spectrum, i.e., in the face of profound hyperimmune response to the infection. Omeprazole (CRank #50, S), used to suppress gastric acid production, alters lysosome acidification and, along with other benzimidazoles, binds to nonstructural protein 3 (nsp3). Blocking this protein, which enhances the virus's ability to evade the immune system [37], was found to interfere in viral formation of SARS-CoV-2[38]. The combination of CRank and strong outcomes highlighted a few other drugs that may be considered for clinical trial based on knowledge of the general pharmacology, including fluvastatin (CRank #199, S), an HMG-CoA reductase inhibitor used to lower cholesterol, but with pleiotropic effects, including anti-inflammatory effects (likely a class effect, as atorvastatin and pitavastatin also had similar effects); ivermectin (CRank #235, S), an anti-parasitic agent; and sildenafil (CRank #493, S), a phosphodiesterase-5 inhibitor.

Taken together, the methodological advances presented here not only suggest potential drug candidates for COVID-19, but offer a principled algorithmic toolset to identify future treatments for diseases underserved by the cost and the timelines of conventional de novo drug discovery processes. As only 918 of the 6,340 drugs prioritized by CRank were screened, a selection driven by compound availability, many potentially efficacious FDA-approved drugs remain to be tested. Finally, it is also possible that some drugs that lacked activity in VeroE6 cells may nevertheless show efficacy in human cells, like loratadine (rank #95, N), which inhibited viral activity in the human cell line Caco-2[39]. Ritonavir, our top-ranked drug, also showed no effect in our screen, despite the fact that over 42 clinical trials are exploring its potential efficacy in patients. In other words, some of the drugs ranked high by CRank may show efficacy, even if they are not among the 77 S&W drugs with positive outcomes.

## Authors Contribution

A.A, D.M.G, M.Z, and X.G performed drug predictions. I.D.V analyzed disease comorbidities. R.D and J.J.P performed the drug experiments and the experimental screens. A.A, D.M.G, I.D.V, M.Z, O.V, X.G, and J.L analyzed the data. J.L manually curated the drug candidates. S.D.G guided A.A with designing diffusion-based similarity implementations. O.V curated list of drugs in clinical trials for COVID-19. A.L.B, D.M.G, I.D.V, and X.G wrote the paper with input from all authors. A.L.B designed the study. All authors read and approved the manuscript.

## Acknowledgments

This work was supported, in part, by NIH grants HG007690, HL108630, and HL119145, and by American Heart Association grants D700382 and CV-19 to J.L; A.L.B is supported by NIH grant 1P01HL132825, American Heart Association grant 151708, and ERC grant 810115-DYNASET. M.Z. is supported, in part, by NSF grants IIS-2030459 and IIS-2033384, and by Harvard Data Science Initiative. J.J.P and R.A.D are supported by NIH grants PO1AI120943, RO1AI128364, RO1Ai125453 and from the Massachusetts Consortium on Pathogen Readiness. We wish to thank Nicolette Lee and Grecia Morales for providing support, Helia Sanchez for helping curate the list of drugs in clinical trials for COVID-19, Marc Santolini for suggestions in the diffusion-based methods and Raj S Dattani for comments on the manuscript.

## Declaration of interests

J.L. and A.L.B are co-scientific founder of Scipher Medicine, Inc., which applies network medicine strategies to biomarker development and personalized drug selection. A.L.B is the founder of Nomix Inc. and Foodome, Inc. that apply data science to health; O.V and D.M.G are scientific consultants for Nomix Inc. I.D.V. is a scientific consultant for Foodome Inc.

# Tables

**Table 1. CRank Predictions for Drug Repurposing.** Top 100 consensus predictions of the drug repurposing pipelines aggregated using the CRank algorithm. The top 100 drugs contain 9 drugs with positive experimental outcomes (S&W), 3 of which are among the top 10 drugs. Drugs highlighted in purple correspond to strong outcomes (S), in orange weak outcomes (W), in green to cytotoxic drugs, while non-highlighted drugs have shown no effect (N) in VeroE6 cells.

- No-effect
- Strong
- Weak
- Cytotoxic

| CRank | Drug Name |
|---|---|
| 1 | Ritonavir |
| 2 | Isoniazid |
| 3 | Troleandomycin |
| 4 | Cilostazol |
| 5 | Chloroquine |
| 6 | Rifabutin |
| 7 | Flutamide |
| 8 | Dexamethasone |
| 9 | Rifaximin |
| 10 | Azelastine |
| 11 | Crizotinib |
| 12 | Urea |
| 13 | Methylprednisolone |
| 14 | Dimethyl sulfoxide |
| 15 | Cortisone acetate |
| 16 | Folic acid |
| 17 | Celecoxib |
| 18 | Betamethasone |
| 19 | Prednisolone |
| 20 | Mifepristone |
| 21 | Budesonide |
| 22 | Prednisone |
| 23 | Oxiconazole |
| 24 | Megestrol acetate |
| 25 | Idelalisib |
| 26 | Econazole |
| 27 | Rabeprazole |
| 28 | Quinine |
| 29 | Ticlopidine |
| 30 | Hydrocortisone |
| 31 | Lansoprazole |
| 32 | Methotrexate |
| 33 | Digoxin |
| 34 | Gefitinib |
| 35 | Enzalutamide |
| 36 | Theophylline |
| 37 | Bicalutamide |
| 38 | Trabectedin |
| 39 | Nelfinavir |
| 40 | Beclomethasone dipropionate |
| 41 | Fluconazole |
| 42 | Aminoglutethimide |
| 43 | Ifosfamide |
| 44 | Hydroxychloroquine |
| 45 | Acetic acid |
| 46 | Cyclophosphamide |
| 47 | Methimazole |
| 48 | Teniposide |
| 49 | Ribavirin |
| 50 | Omeprazole |
| 51 | Chlorambucil |
| 52 | Citalopram |
| 53 | Bortezomib |
| 54 | Leflunomide |
| 55 | Dimethyl fumarate |
| 56 | Teriflunomide |
| 57 | Colchicine |
| 58 | Phenylbutyric acid |
| 59 | Progesterone |
| 60 | Triamcinolone |
| 61 | Medroxyprogesterone acetate |
| 62 | Tioguanine |
| 63 | Quercetin |
| 64 | Clobetasol |
| 65 | Letrozole |
| 66 | Etoposide |
| 67 | Mebendazole |
| 68 | Adenosine |
| 69 | Mesalazine |
| 70 | Nevirapine |
| 71 | Belinostat |
| 72 | Mitomycin |
| 73 | Malathion |
| 74 | Ixekizumab |
| 75 | Vindesine |
| 76 | Secukinumab |
| 77 | Rifapentine |
| 78 | Bilastine |
| 79 | Clotrimazole |
| 80 | Erlotinib |
| 81 | Panobinostat |
| 82 | Warfarin |
| 83 | Busulfan |
| 84 | Goserelin |
| 85 | Hydroxyurea |
| 86 | Temsirolimus |
| 87 | Abiraterone |
| 88 | Miconazole |
| 89 | Ketorolac |
| 90 | Exemestane |
| 91 | Oxymetholone |
| 92 | Pentamidine |
| 93 | Diclofenac |
| 94 | Aminophylline |
| 95 | Loratadine |
| 96 | Fexofenadine |
| 97 | Terbinafine |
| 98 | Verapamil |
| 99 | Clopidogrel |
| 100 | Rivaroxaban |

**Table** 2 **Drugs with Positive Experimental Outcomes.** List of the 77 drugs with a positive outcome (S&W) from in vitro screen[17]. Drug response classification was obtained by a two-step model for drug response (see SI Section 2.3). Purple drugs show strong effect (S), and orange drugs showed weak effect (W).

Strong
Weak

| CRank | Drug Name | CRank | Drug Name | CRank | Drug Name |
|---|---|---|---|---|---|
| 5 | Chloroquine | 423 | Pitavastatin | 742 | Mianserin |
| 6 | Rifabutin | 431 | Tenoxicam | 755 | Clofazimine |
| 9 | Rifaximin | 438 | Quinidine | 767 | Chlorpromazine |
| 10 | Azelastine | 456 | Sertraline | 772 | Imipramine |
| 16 | Folic acid | 460 | Ingenol mebutate | 830 | Promazine |
| 32 | Methotrexate | 463 | Norelgestromin | 900 | L-Alanine |
| 33 | Digoxin | 493 | Sildenafil | 917 | Moxifloxacin |
| 44 | Hydroxychloroquine | 499 | Eliglustat | 933 | Tasimelteon |
| 50 | Omeprazole | 518 | Ulipristal | 995 | Vandetanib |
| 113 | Clobetasol propionate | 553 | Cinacalcet | 1000 | Azilsartan medoxomil |
| 118 | Auranofin | 556 | Perphenazine | 1020 | Frovatriptan |
| 120 | Vinblastine | 558 | Idarubicin | 1034 | Zolmitriptan |
| 199 | Fluvastatin | 564 | Perhexiline | 1035 | Procarbazine |
| 210 | Clomifene | 569 | Amiodarone | 1093 | Asenapine |
| 233 | Ibuprofen | 577 | Duloxetine | 1107 | Dyclonine |
| 235 | Ivermectin | 585 | Toremifene | 1140.5 | Clemastine |
| 243 | Atorvastatin | 586 | Afatinib | 1194 | Prochlorperazine |
| 253 | Pralatrexate | 601 | Amitriptyline | 1222 | Miglustat |
| 263 | Cobimetinib | 626 | Meclizine | 1224 | Prenylamine |
| 269 | Hydralazine | 635 | Valsartan | 1276 | Dalfampridine |
| 297 | Propranolol | 651 | Eletriptan | 1314 | Cinchocaine |
| 317 | Osimertinib | 673 | Sotalol | 1355 | Methotrimeprazine |
| 348 | Vincristine | 678 | Thioridazine | 1396 | Methylthioninium |
| 367 | Doxazosin | 695 | Chlorcyclizine | 1403 | Metixene |
| 397 | Rosiglitazone | 707 | Omacetaxine mepesuccinate | 1443 | Trifluoperazine |
| 398 | Aminolevulinic acid | 721 | Candesartan | | |

# Figures

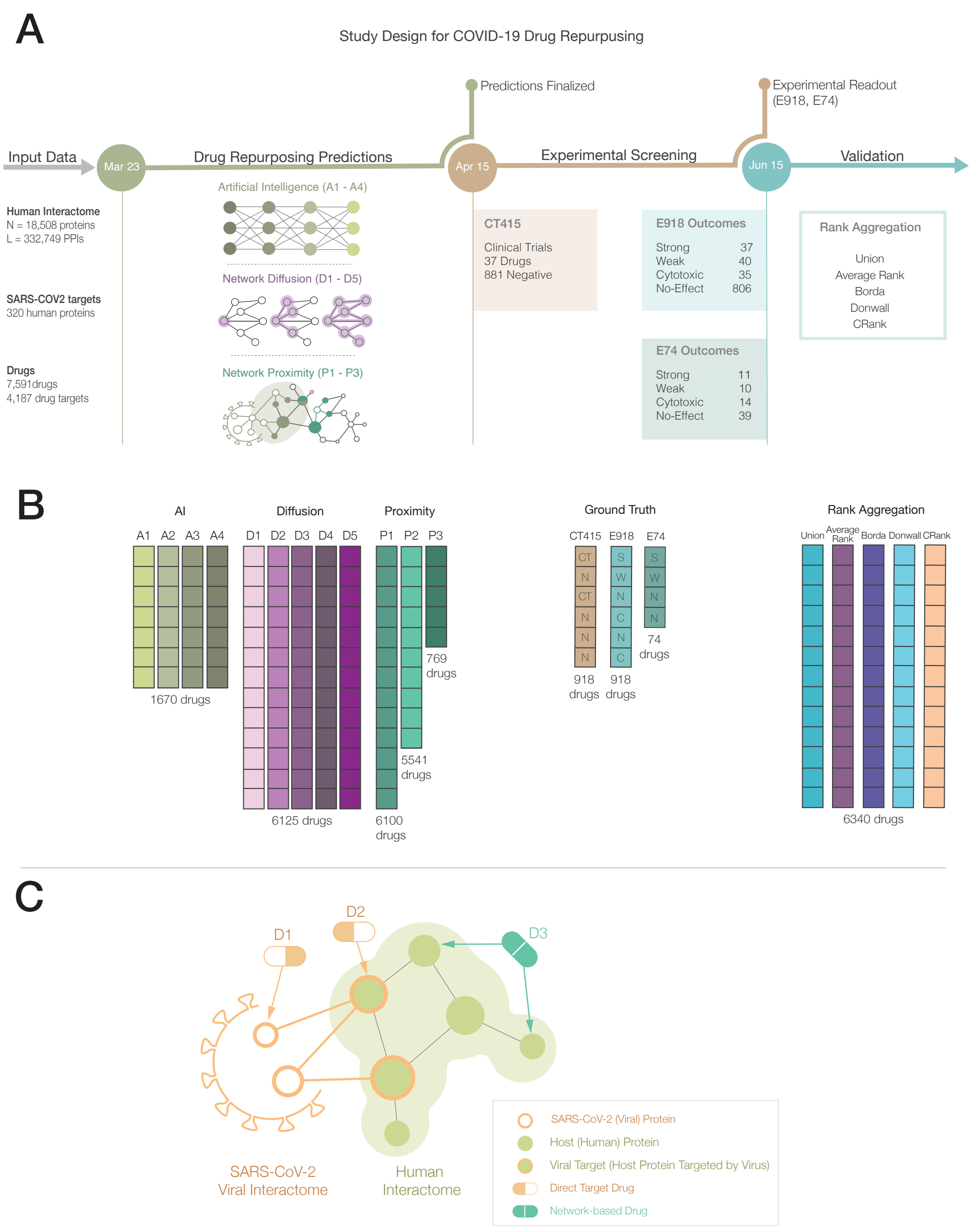

**Figure 1 Network Medicine Framework for Drug Repurposing. (A)** Study Design and Timeline. Following the publication of host-pathogen protein-protein interactions[20] – March, 23rd, 2020 – we implemented three drug repurposing algorithms, relying on AI (A1-A4), network diffusion (D1-D5) and proximity (P1-P3), together resulting in 12 predictive ranking lists (pipelines, shown in **(B)**). Each pipeline offers predictions for a different number of drugs, what were frozen on April 15, 2020. We then identified 918 drugs for which all pipelines but P3 offered predictions, and experimentally validated their effect on the virus in VeroE6 cells[17]. The experimental (E918, E74) and clinical trial lists C415 offered the ground truth for validation and rank aggregation. **(C)** Direct target drugs bind either to a viral protein (D1) or to a host protein target of the viral proteins (D2). Network drugs (D3), in contrast, bind to the host proteins and limit viral activity by perturbing the host subcellular network.

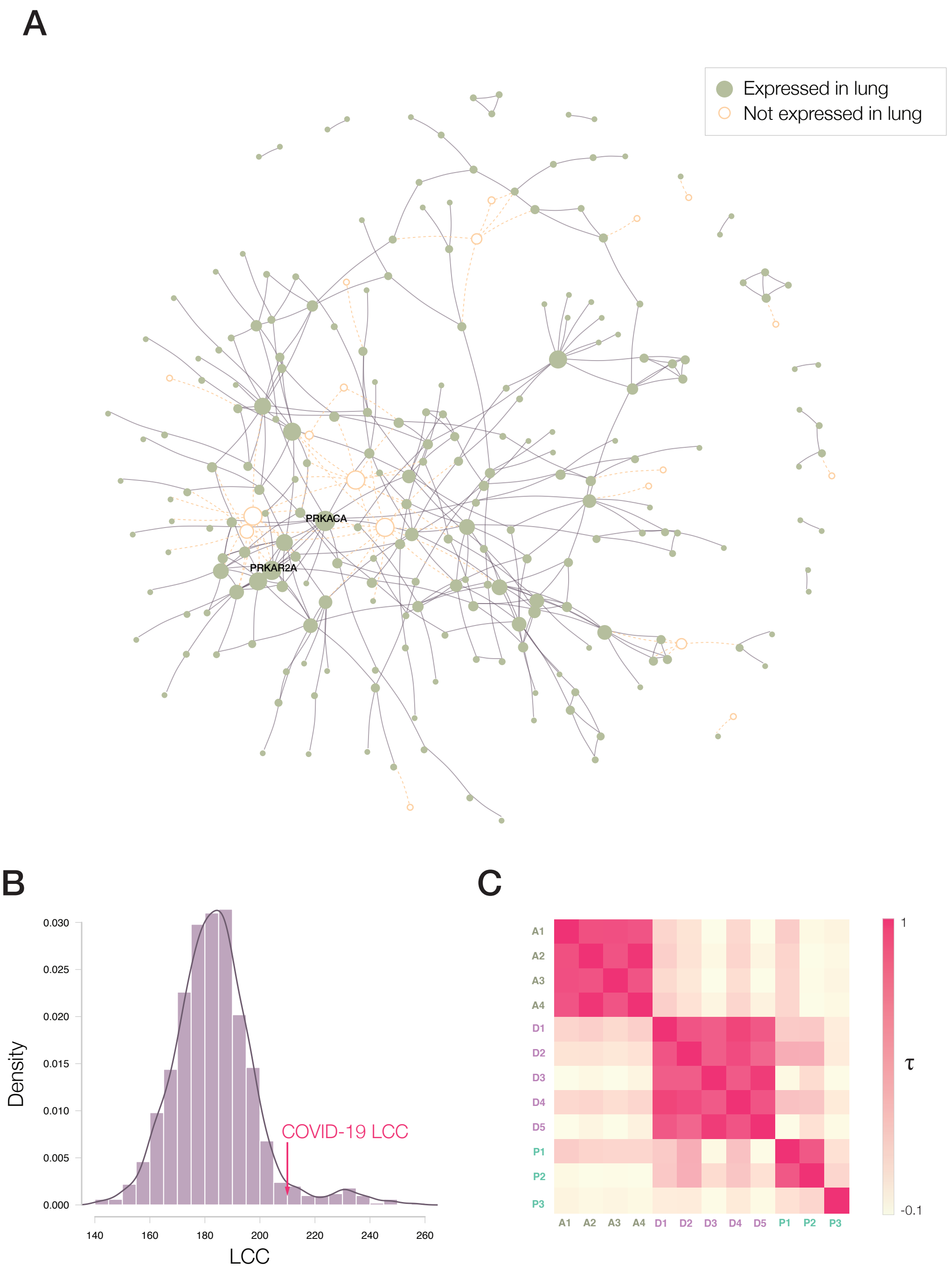

**Figure 2. COVID-19 Disease Module. (A)** Proteins targeted by SARS-CoV-2 are not distributed randomly in the human interactome, but form a large connected component (LCC) consisting of 208 proteins, and multiple small subgraphs, shown in the figure. Almost all proteins in SARS-CoV-2 LCC are also expressed in the lung tissue, potentially explaining the effectiveness of the virus in causing pulmonary manifestations of the disease. **(B)** The random expectation of the LCC size indicates that the observed COVID-19 LCC, whose size is indicated by the red arrow, is larger than expected by chance (Z-score=1.65). **(C)** Heatmap of the Kendall $\tau$ statistic showing that the ranking list predicted by the different methods (A,D,P) are not correlated. We observe, however high correlations among the individual ranking list predicted by the same predictive method.

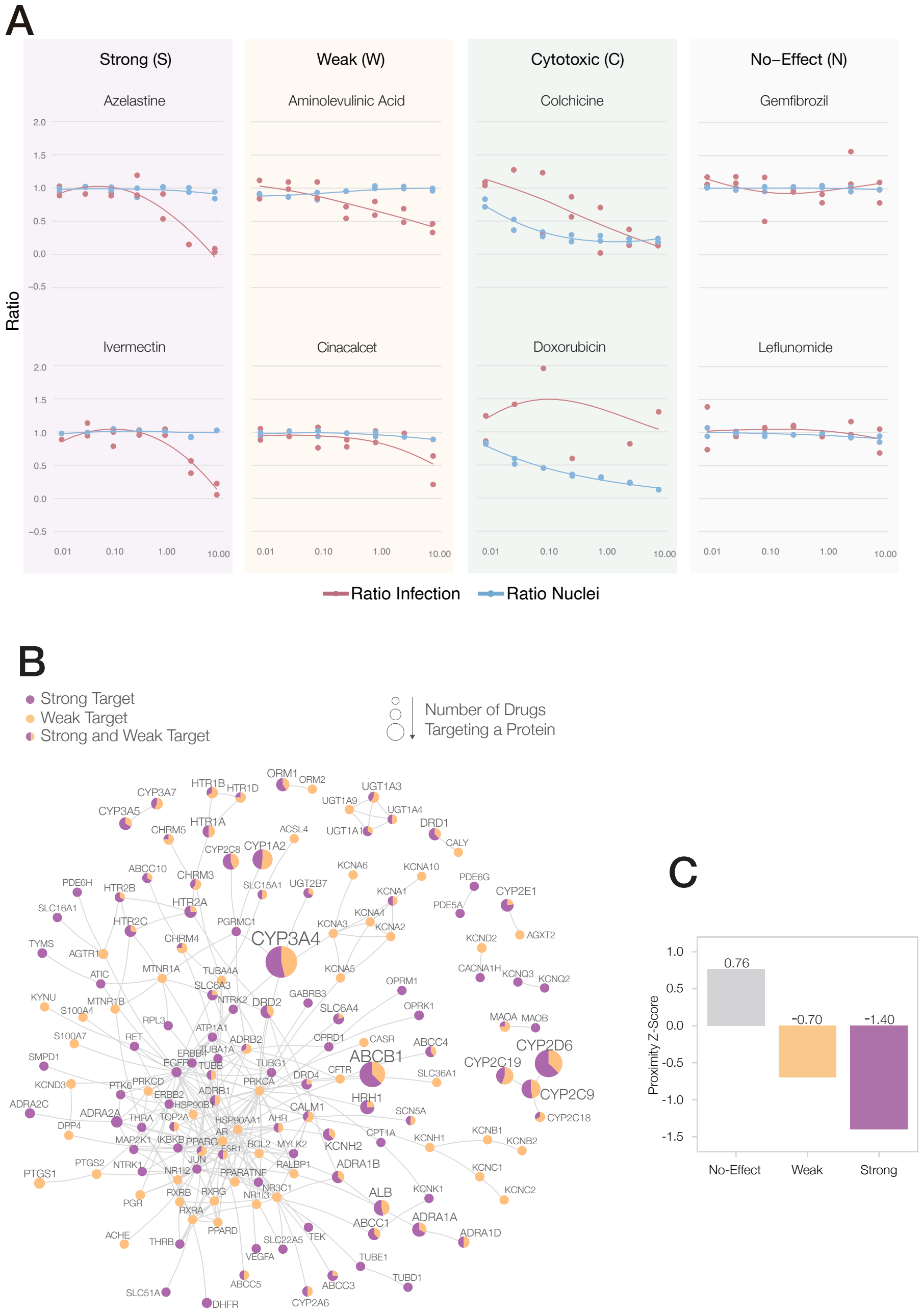

**Figure 3 Experimental Outcomes and Network Origins. (A)** Examples of dose-response curves for eight of the 918 experimentally validated drugs[17], illustrating the four observed outcomes (S, W, C and N). VeroE6 cells were challenged in vitro with SARS-CoV-2 virus and treated with the drug over a range of doses (from 8 nM to 8 μM). A two-steps drug-response model (see SI Section 2.3) was used to classify each drug into Strong, Weak, Cytotoxic or No-Effect categories, according to their response to the drug in different doses and cell and viral reduction. **(B)** The sub-network formed by the targets of the 77 S&W drugs within the interactome. The link corresponds to binding interactions. Purple proteins are targeted by S drugs only; orange by W drugs only; proteins targeted by both S&W drugs are shown as pie charts, proportional to the number of targets in each category. **(C)** The targets of N drugs have a positive proximity Z-Score to the COVID-19 module, meaning they are further from the COVID-19 module than random expectation. By contrast, the targets of S&W drugs are more proximal (closer) to the COVID-19 module than expected by change, suggesting that their COVID-19 vicinity contribute to their ability to alter the virus's ability to infect the cells.

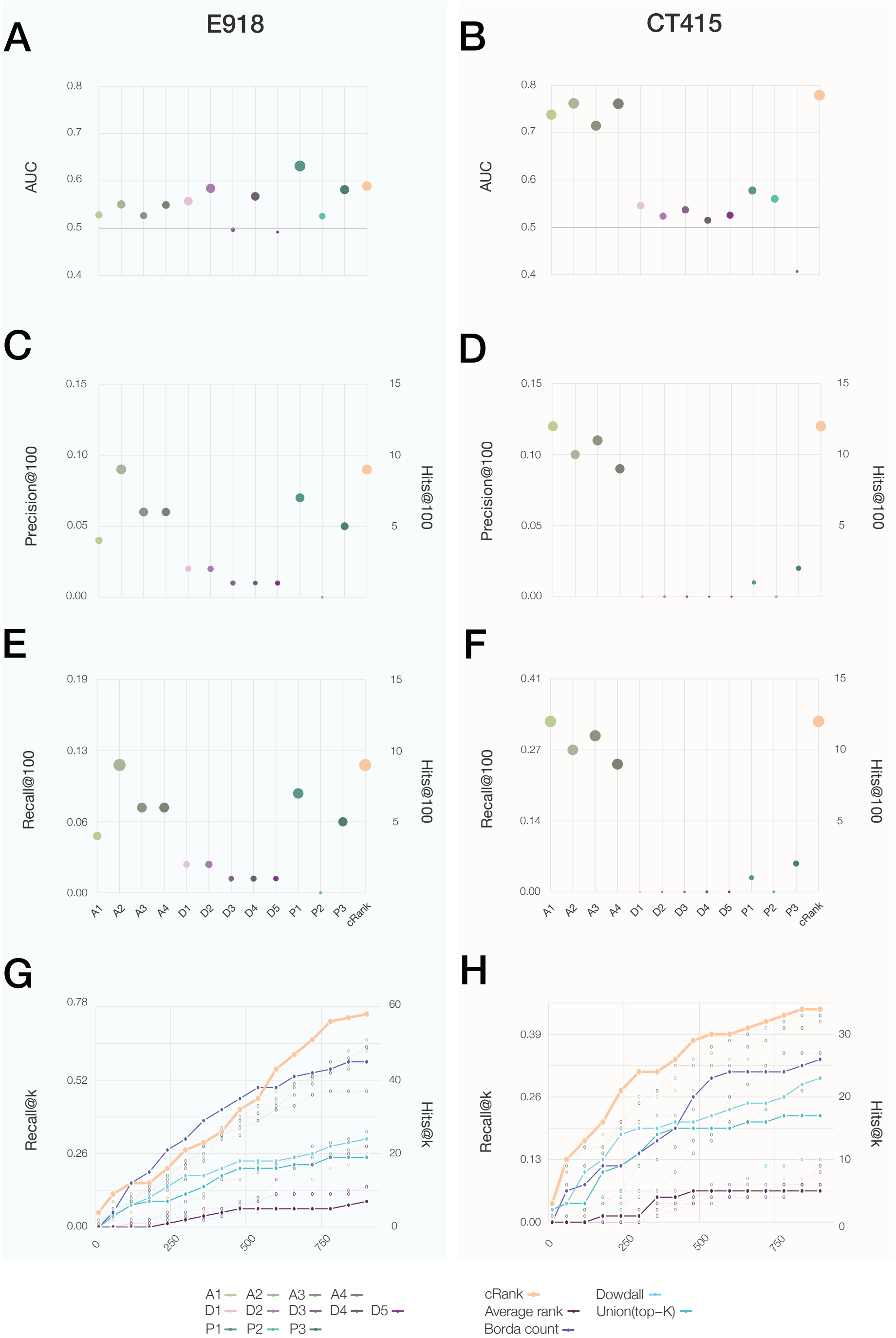

**Figure 4 Performance of the Predictive Pipelines. (A,B)** AUC (Area under the Curve), **(C,D)** Precision at 100, and **(E,F)** Recall at 100, for twelve pipelines tested for drug repurposing, each plot using as a gold standard the S&W drugs in E918 (left column) and drugs under clinical trials for treating COVID-19 as of April 15th, 2020 (CT415, right column). **(G,H)** The top $K$ precision and recall for the different rank aggregation methods (connected points), compared to the individual pipelines (empty symbols) documenting the consistent predictive performance of CRank. Similar results are shown for two other datasets in Figure S8: the prospective expert curated E74 and the clinical trial data refreshed on 06/15/20 (CT615)

# Supplementary Information: Network Medicine Framework for Identifying Drug Repurposing Opportunities for COVID-19

Deisy Morselli Gysi, Ítalo Do Valle, Marinka Zitnik, Asher Ameli, Xiao Gan, Onur Varol, Susan Dina Ghiassian, JJ Patten, Robert Davey, Joseph Loscalzo, and Albert-László Barabási

## Table of Contents

# 1 Network-Based Drug Repurposing For COVID-19

## 1.1 Human Interactome and SARS-CoV-2 and Drug Targets

The human interactome was assembled from 21 public databases that compile experimentally derived protein-protein interaction (PPI) data: 1) binary PPIs, derived from high-throughput yeast-two hybrid (Y2H) experiments (HI-Union[1]), three-dimensional (3D) protein structures (Interactome3D[2], Instruct[3], Insider[4]), or literature curation (PINA[5], MINT[6] , LitBM17[1], Interactome3D, Instruct, Insider, BioGrid[7], HINT[8], HIPPIE[9], APID[10], InWeb[11]); 2) PPIs identified by affinity purification followed by mass spectrometry present in BioPlex[12], QUBIC[13], CoFrac[14], HINT, HIPPIE, APID, LitBM17, and InWeb; 3) kinase substrate interactions from KinomeNetworkX[15] and PhosphoSitePlus[16]; 4) signaling interactions from SignaLink[17] and InnateDB[18]; and 5) regulatory interactions derived by the ENCODE consortium. We used the curated list of PSI-MI IDs provided by Alonso-Lopez et. al. (2019)[10] for differentiating binary interactions among the several experimental methods present in the literature-curated databases. For InWeb, interactions with curation scores < 0.175 (75th percentile) were not considered. All proteins were mapped to their corresponding Entrez ID (NCBI) and the proteins that could not be mapped were removed. The final interactome used in our study contains 18,505 proteins, and 327,924 interactions (Table S3). We retrieved interactions between 26 SARS-CoV-2 proteins and 332 human proteins reported by Gordon, et. al. (2020)[19] (Table S4). We retrieved drug target information from the DrugBank database[20], which contains 24,609 interactions between 6,228 drugs and their 3,903 targets, and drug target interaction data curated from the literature for 25 drugs (Table S5). We also obtained from the DrugBank database differentially expressed genes (DEGs) identified by exposure of drugs to different cell lines (Table S6). The Largest Connected Component (LCC) of human proteins that bind to SARS-CoV-2 proteins was calculated using a degree-preserving approach[21], which prevents the repeated selection of the same high degree nodes, setting 100 degree bins in 1,000 realizations.

## 1.2 Lung Gene Expression (Fig 2A)

We evaluated gene expression in the lung by using the GTEX database[22], considering genes with a median count lower than 5 transcripts (raw counts) as not expressed.

## 1.3 Disease Comorbidities

Pre-existing conditions worsen prognosis and recovery of COVID-19 patients[23]. Previous work showed that the disease relevance of human proteins targeted by a virus can predict the signs, symptoms, and diseases caused by that pathogen[24]. This prompted us to identify diseases whose molecular mechanisms overlap with cellular processes targeted by SARS-CoV-2, allowing us to predict potential comorbidity patterns[25–27]. We retrieved 3,173 disease-associated genes for 299 diseases[28], finding that 110 of the 332 proteins targeted by SARS-CoV-2 are implicated in other human diseases; however, the overlap between SARS-CoV-2 targets and the pool of the disease-associated genes was not statistically significant (Fisher's exact test; FDR-BH $p_{adj}$ -value> 0.05). We evaluated the network-based overlap between the proteins associated with each of the 299 diseases and the host protein targets of SARS-CoV-2 using the $S_{vb}$ metric[28], where $S_{vb} < 0$ signals a network-based overlap between the SARS-CoV-2 viral targets $v$ and the gene pool associated with disease $b$. We found that $S_{vb} > 0$ for each disease, indicating that COVID-19 disease module does not directly overlap with any major disease module (*Figure S1* and Table S7). The diseases closest to the COVID-19 disease module (smallest $S_{vb}$) included several cardiovascular diseases and cancers, whose comorbidity in COVID-19 patients is well documented[29–31] (*Figure S2*). The same metric predicted comorbidity with neurological diseases, in line with our observation that the host protein targets are expressed in the brain (Table S7).

In summary, we found that the SARS-CoV-2 host protein targets do not overlap with proteins associated with any major diseases, indicating that a potential COVID-19 treatment cannot be derived from the arsenal of therapies approved for a specific disease. These findings argue for a strategy that maps drug targets without regard to their localization within a particular disease module. However, the disease modules closest to the SARS-CoV-2 viral targets are those with noted comorbidity for COVID-19 infection, such as pulmonary and cardiovascular diseases. We also found multiple network-based evidences linking the virus to the nervous system, a less explored comorbidity, consistent with the observations that many infected patients initially lose olfactory function and taste[32], and 36% of patients with severe infection who require hospitalization have neurological manifestations.

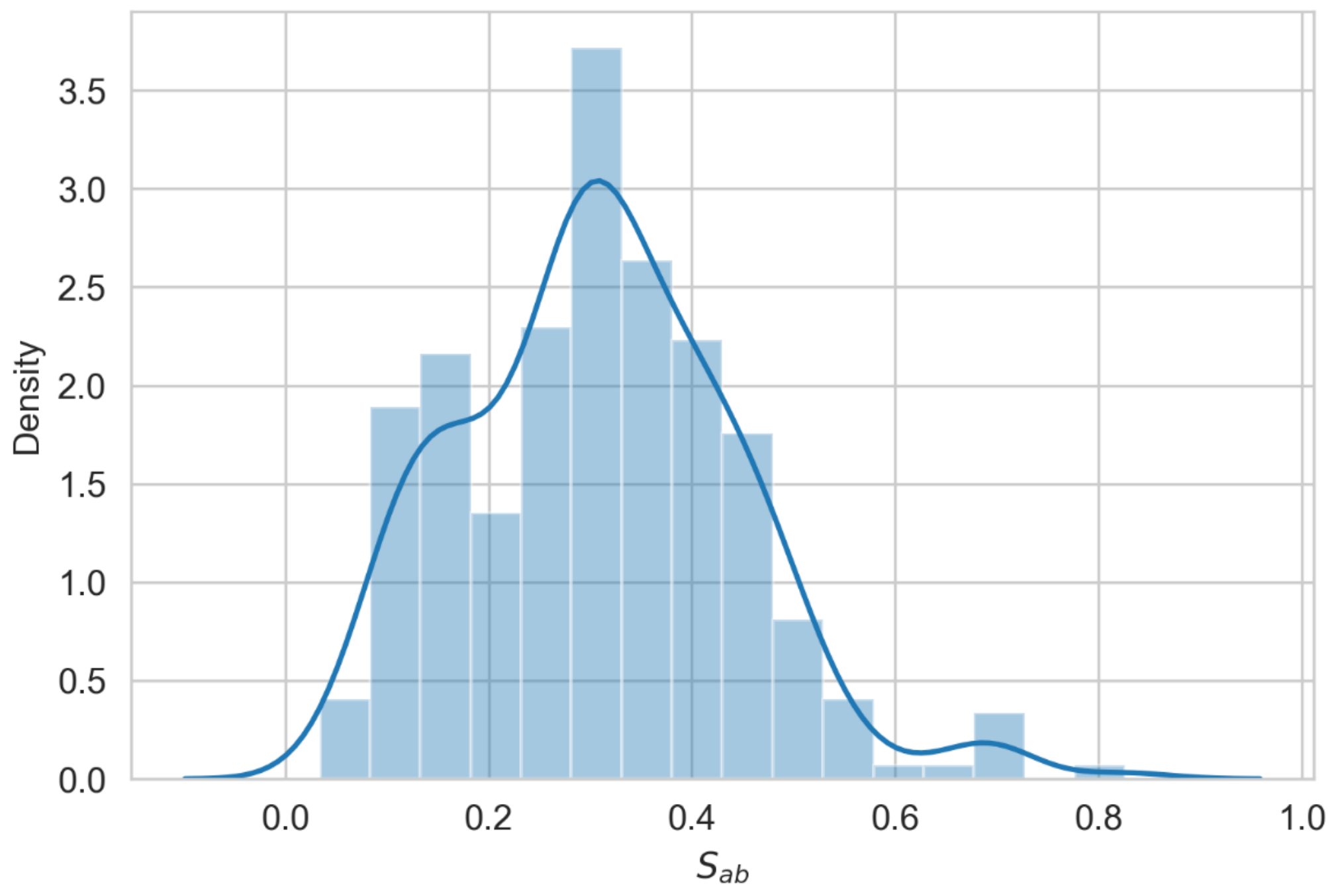

**Figure S1 - Distribution of the Network Overlap Measure $S_{vb}$ Between 299 Diseases and COVID-19 Targets.** $S_{vb}$ values represent the network-based overlap between SARS-COV2 targets *v* and the genes associated with each disease *b.*

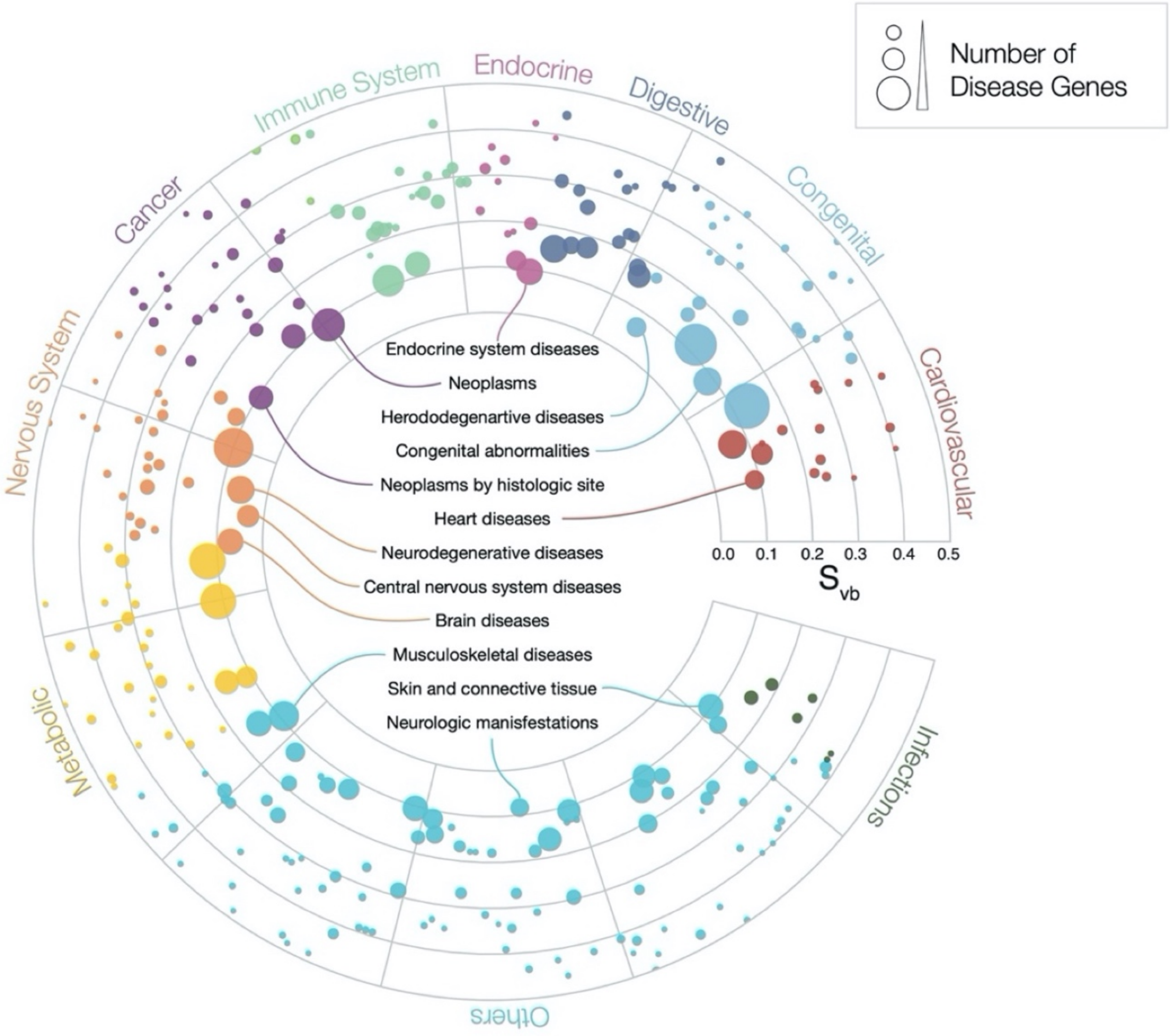

**Figure S2 - Disease Comorbidity Measured by the Network Overlap Between COVID-19 Targets and 299 Diseases.** The figure represents each disease as a circle whose radius reflects the number of disease genes associated with it[28]. The diseases closest to the center, whose names are marked, are expected to have higher comorbidity with the COVID-19 outcome. The farther a disease is from the center, the more distant are its disease proteins from the COVID-19 viral targets. Disease Comorbidity. We measured the network proximity between COVID-19 targets and 299 diseases. The figure represents each disease as a circle whose radius reflects the number of disease genes associated with it[28]. The diseases closest to the center, whose names are marked, are expected to have higher comorbidity with the COVID-19 outcome. The farther a disease is from the center, the more distant are its disease proteins from the COVID-19 viral targets

## 1.4 Drug Repurposing Prediction Algorithms

### 1.4.1 Artificial Intelligence Based Algorithm (A1-A4)

We designed a graph neural network for COVID-19 treatment recommendations based on a previously developed graph neural network (GNN) architecture[33] (Figure S3).The multimodal graph is a heterogeneous graph $G = (V, R)$ with $N$ nodes $v_i \in V$ representing three distinct types of biomedical entities (i.e., drugs, proteins, diseases), and triplets, i.e., labeled edges $(v_i, r, v_j) \in R$ representing four semantically distinct types of edges $r$ between the entities (i.e., protein-protein interactions, drug-target associations, disease-protein associations, and drug-disease indications).

**COVID-19 drug treatment recommendation task.** We cast the COVID-19 treatment recommendation task as a link prediction problem on the multimodal graph. The task was to predict new edges between drug and disease nodes such that a predicted link between a drug node $v_i$ and a disease node $v_j$ should indicate that drug $v_i$ is a promising treatment for disease $v_j$ (e.g., COVID-19). Our graph neural network is an end-to-end trainable model for link prediction on the multimodal graph and has two main components: (1) An encoder: a graph convolutional network operating on $G$ and producing embeddings for nodes in $G$; and (2) A decoder: a model optimizing embeddings such that they are predictive of approved drug indications.

**Overview of graph neural architecture.** The neural message passing encoder took as input a graph $G$ and produced a node d-dimensional embedding $z_i \in R^d$ for every drug and disease node in the graph. We used the encoder[33] that learned a message-passing algorithm[34] and aggregation procedure to compute a function of the entire graph that transformed and propagated information across graph $G$[34]. The graph convolutional operator took into account the first-order neighborhood of a node and applied the same transformation across all locations in the graph. Successive application of these operations then effectively convolved information across the $K^{th}$ order neighborhood (i.e., embedding of a node depends on all the nodes that are at most $K$ steps away), where $K$ is the number of successive operations of convolutional layers in the neural network model. The graph convolutional operator takes the form

$$\boldsymbol{h}_i^{(k+1)} = \phi\left(\sum_r \sum_{j\in N_r^i} \alpha_r^{ij}\, \boldsymbol{W}_r^{(k)} \boldsymbol{h}_j^{(k)} + \alpha_r^i \boldsymbol{h}_i^{(k)}\right), \quad (1)$$

where $\boldsymbol{h}_i^{(k)} \in R^{d(k)}$ is the hidden state of node $v_i$ in the $k^{th}$ layer of the neural network with $d(k)$ being the dimensionality of this layer's representation, $r$ is an edge type, matrix $\boldsymbol{W}_r^{(k)}$ is an edge-type specific parameter matrix, $\phi$ denotes a non-linear element-wise activation function (*i.e.*, a rectified linear unit), and $\alpha_r$ denote attention coefficients[35]. To arrive at the final embedding $z_i \in R^d$ of node $v_i$, we compute its representation as $z_i = \boldsymbol{h}_i^{(k)}$. Next, the decoder takes node embeddings and combines them to reconstruct labeled edges in $G$. In particular, the decoder scores a $(v_i, r, v_j)$ triplet through a function $g$ whose goal is to assign a score $g(v_i, r, v_j)$ representing how likely it is that drugs $v_i$ will treat disease $v_j$ (i.e., $r$ denotes an 'indication' relationship)[35].

**Training the graph neural network.** During model training, we optimized model parameters using the max-margin loss functions to encourage the model to assign higher probabilities to successful drug indications $(v_i, r, v_j)$ than to random drug-disease pairs. We took an end-to-end optimization approach that jointly optimized over all trainable parameters and propagated loss function gradients through both the encoder and the decoder. To optimize the model, we trained it for a maximum of 100 epochs (training iterations) using the Adam optimizer[36] with a learning rate of 0.001. We initialized weights using the initialization described in[37]. To make the model comparable to other drug repurposing methodologies in this study, we did not integrate additional side information into node feature vectors; instead, we used one-shot indicator vectors[38] as node features. For the model to generalize well to unobserved edges, we applied a regular dropout[39] to hidden layer units (Eq. (10)). In practice, we used efficient sparse matrix multiplications, with complexity linear in the number of edges in $G$, to implement the model. We used a 2-layer neural architecture with $d_1 = 32$, $d_2 = 32$, $d_i = 128$ hidden units in input, output, and intermediate layer, respectively; a dropout rate of 0.1; and a max-margin of 0.1. We used mini-batching[40] by sampling triplets R from the multimodal graph $G$. That is, we processed multiple training mini-batches (mini-batches are of size 512), each obtained by sampling only a fixed number of triplets, resulting in dynamic batches that changed during model training.

**Constructing ranked lists of candidate drugs for COVID-19.** We generated four lists of candidate drugs for COVID-19. To generate the lists, we used embeddings returned by the graph neural network, in particular, embeddings learned for nodes representing either COVID-19 or drugs in multimodal graph $G$. The embedding vectors for diseases and drugs are provided in Table S8 and Table S9, respectively. The pipeline A1 searches for drugs that are in the vicinity of the COVID-19 disease by calculating the cosine distance between COVID-19 and all drugs in the decoded embedding space[41]. The decoding is based on the $N = 10$ nearest neighboring nodes in the embedding space, with a minimum distance between nodes of $D = 0.25$. The pipeline A2 prevents that nodes in the decoding embedding space from packing together too closely, by using $D = 0.8$ and keeping $N$ unchanged. These constraints push the structures apart into softer, more general features, offering a better overarching view of the embedding space at the loss of the more detailed structure. Pipeline A3 forces the decoding to concentrate on the very local structure by using N = 5, to explore a smaller neighborhood, while setting the minimum distance at a midrange point of D = 0.5. Pipeline A4 focuses on a broader view of the embedding space by setting N=10 and D = 1. Finally, to obtain lists of candidate drugs, each pipeline ranked drugs based on the pipeline-defined distances of drugs to COVID-19 (Figure S3). Intuitively, parameter $N$ constrained the size of the local neighborhood each pipeline looked at in the embedding space when calculating the distances, and parameter $D$ controlled how tightly the pipeline was allowed to pack the embeddings together.

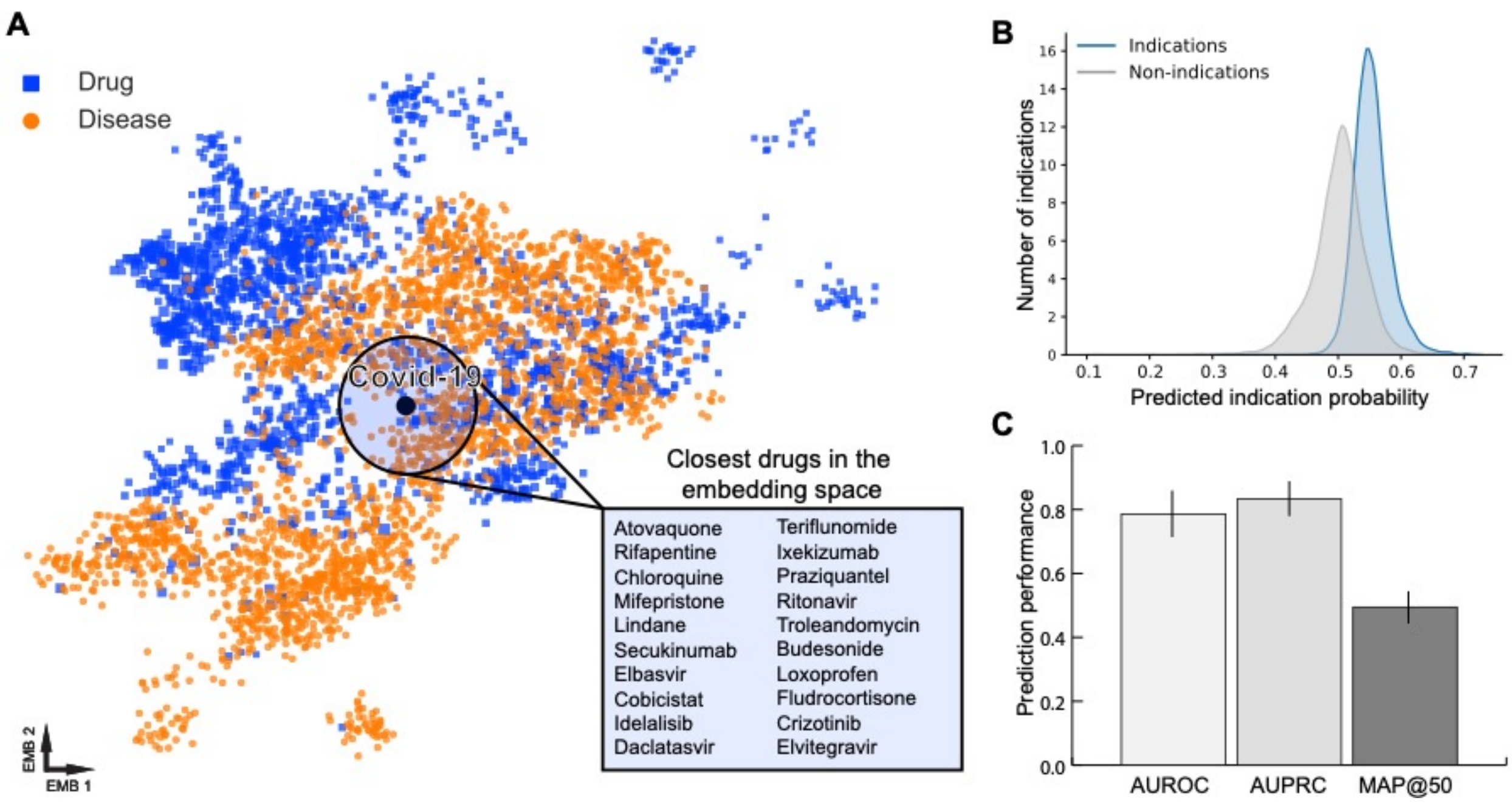

Figure S3 - Overview of AI-based Strategy for Drug Repurposing. (A) Visualization of the learned embedding space. Every point represents a drug (in blue) or a disease (in orange). If a drug and a disease are embedded close together in this space, this means the local interaction neighborhoods of the drug and the disease in the multimodal graph are predictive of whether the drug can treat the disease. (B) Testing of the graph neural network (GNN) model. Probability distributions of approved indications and non-indications (i.e., random drug-disease pairs) learned by the GNN model are well-separated, indicating the model can distinguish between true indications and random drug-disease pairs. (C) Predictive performance of the GNN model on the test set of drug indications that were held-out during model training. Higher values indicated better performance (AUROC, Area under the ROC curve; AUPRC, Area under the PR curve; MAP@50, Mean average precision at top 50).

### 1.4.2 Diffusion-Based Algorithms (D1-D5)

The diffusion state distance (DSD)[42] algorithm uses a graph diffusion property to derive a similarity metric for pairs of nodes that takes into account how similarly they affect the rest of the network. We calculate the expected number of times $He(A, B)$ that a random walker starting at node $A$ visits node $B$, representing each node by the vector[42]:

$$He\ (V_i) = [He(V_i, V_1), He(V_i, V_2), He(V_i, V_3), \ldots, He(V_i, V_n)], \qquad (2)$$

which describes how a perturbation initiated from that node affects other nodes in the interactome. The similarity between nodes $A$ and $B$ is provided by the L1-norm of their corresponding vector representations:

$$DSD(A, B) = ||He(A) - He(B)||. \qquad (3)$$

Inspired by the DSD, we developed five new metrics to calculate the impact of drug targets $T$ on the SARS-CoV-2 targets $V$. The first (Pipeline D1) is defined as:

$$I_{DSD}^{min} = \frac{1}{||T||} \sum_{t \in T} \min_{v \in V} DSD\ (t, v), \qquad (4)$$

where $DSD(s, t)$ represents the diffusion state distance between nodes $t$ and $v$. Since the L1-norm of two large vectors may result in loss of information[43], we also used the metrics (Pipeline D2):

$$I_{KL}^{min} = \frac{1}{||T||} \sum_{t \in T} \min_{v \in V} KL\ (t, v) \qquad (5)$$

and (Pipeline D3):

$$I_{KL}^{med} = \frac{1}{||T||} \sum_{t \in T} \operatorname{median}_{v \in V} KL\ (t, v), \qquad (6)$$

where $KL$ is the Kullback-Leibler ($KL$) divergence between the vector representations of the nodes $t$ and $s$. Finally, to provide symmetric measures, we tested the metrics (Pipeline D4):

$$I_{JS}^{min} = \frac{1}{||T||} \sum_{t \in T} \min_{v \in V} JS\ (t, v) \qquad (7)$$

and (Pipeline D5)

$$I_{JS}^{med} = \frac{1}{||T||} \sum_{t \in T} \operatorname{median}_{v \in V} JS\ (t, v). \qquad (8)$$

where $JS$ is the Jensen Shannon ($JS$) divergence between the vector representations of nodes $t$ and $s$. All five measures assume $t \neq s$.

### 1.4.3 Proximity Algorithm (P1-P3)

Given $V$, the set of COVID-19 virus targets, $T$, the set of drug targets, and $d(v,t)$, the shortest path length between nodes $v \in V$ and $t \in T$ in the network, we define[21]:

$$d_c(V,T) = \frac{1}{||T||} \sum_{t \in T} \min_{v \in V} d(v,t). \quad (9)$$

We determined the expected distances between two randomly selected groups of proteins, matching the size and degrees of the original *V* and *T* sets. To avoid repeatedly selecting the same high degree nodes, we use degree-binning[21]. The mean $\mu_{d(V,T)}$ and standard deviation $\sigma_{d(V,T)}$ of the reference distribution allows us to convert the absolute distance $d_c$ to a relative distance $Z_{dc}$, defined as:

$$Z_{d_c} = \frac{d_c - \mu_{d_c}(V,T)}{\sigma_{d_c}(V,T)}. \quad (10)$$

We implemented three versions of the proximity algorithm: 1) relying on all drug targets (P1); 2) ignoring drug targets identified as drug carriers, transporters, and drug-metabolizing enzymes (P2); and 3) on differentially expressed genes (DEGs) identified by exposure of each drug to cultured cells, which was obtained from DrugBank's compilation of 17,222 DEGs linked to 793 drugs in multiple cell lines (Table S6).

## 1.5 Network Properties of Prediction Algorithms

### 1.5.1 Explanatory Subgraphs

For each pipeline, we identified "explanatory subgraphs" to help understand the predictions made by the respective pipeline. The key idea was to summarize where in the data the pipeline looks for evidence for their predictions. Given a particular prediction, an explanatory subgraph is a small sub-network of the entire network considered by the pipeline that is most influential for the prediction and contributes most to the predictive power. For the proximity method (P), the explanatory subgraphs can be derived exactly, representing the set of nodes contributing to proximity. For the artificial intelligence-based methods (A), the subgraphs were extracted using a GNN Explainer algorithm[44]. GNNExplainer specifies an explanation as a subgraph of the entire network the GNN was trained on, such that the subgraph maximizes the mutual information with the GNN's prediction. This is achieved by formulating a mean field variational approximation and learning a real-valued graph mask, which selects the important subgraph using counterfactual reasoning. For the diffusion method, we first identified the SARS-CoV-2 targets (seeds) that have the maximum (or median, depending on the pipeline) similarity with the drug targets under consideration. Once the seeds are identified for each drug target, we extract the vector representation of the target and the corresponding seeds. Each element of these vectors corresponds to a node in the network:

$$t: [r_1, r_2, r_3, \dots, r_n]$$

$$s: [w_1, w_2, w_3, \dots, w_n]$$

Each pipeline performs an element-wise comparison of these two vectors to calculate similarity values, defined as similarity terms, using:

$$term_i^{DSD}(t, s) = |r_i - w_i| \qquad (11)$$

$$term_i^{KL}(t, s) = r_i \log\left(\frac{r_i}{w_i}\right) \qquad (12)$$

$$term_i^{JS}(t,s) = \frac{1}{2}[r_i \log\left(\frac{r_i}{m_i}\right) + w_i \log(\frac{w_i}{m_i})],\ m_i = \frac{r_i + w_i}{2} \quad (13)$$

These distance similarity terms collectively contribute to each drug's ranking score. Among all 18,446 nodes, we are only interested in those whose variations lead to the current ranking (drug prediction scores). Therefore, we applied a feature selection algorithm to eliminate the network nodes (features) that do not contribute to the predicted scores (outcomes). This task is done by training a regression tree model (DecisionTreeRegressor model, from Python 3 scikit-learn package) where feature values are the similarity terms (as defined above) between drug targets and the corresponding seeds. This resulted in 2,507 important features for pipeline D1 (DSD-min), 2198 for D2 (KL-min), 2,263 for D3 (KL-med), 1,655 for D4 (JS-min), and 1,817 for D5 (JS-med). Important features are those with non-zero importance value as characterized by the Regressor model.

Once the important features/nodes are extracted, we search this space to identify the explanatory network of each set of drug targets. To do so, we rank the similarity terms of each target and the corresponding seeds on the space of important features and identify the nodes with the highest contribution to the similarity measure such that they satisfy the following equation:

$$\log_{10}\left(\frac{l}{term_i}\right) \leq 1,\ l = \max(term_i),\ i \in V \quad (14)$$

If a drug has multiple targets or if each target has multiple corresponding seeds (seeds with the same similarity to a target), the results are aggregated. The explanatory network of a target that happens to be a seed is that seed itself.

**Figure S*4*** shows the similarities and differences among the explanatory subgraphs of the different prediction pipelines.

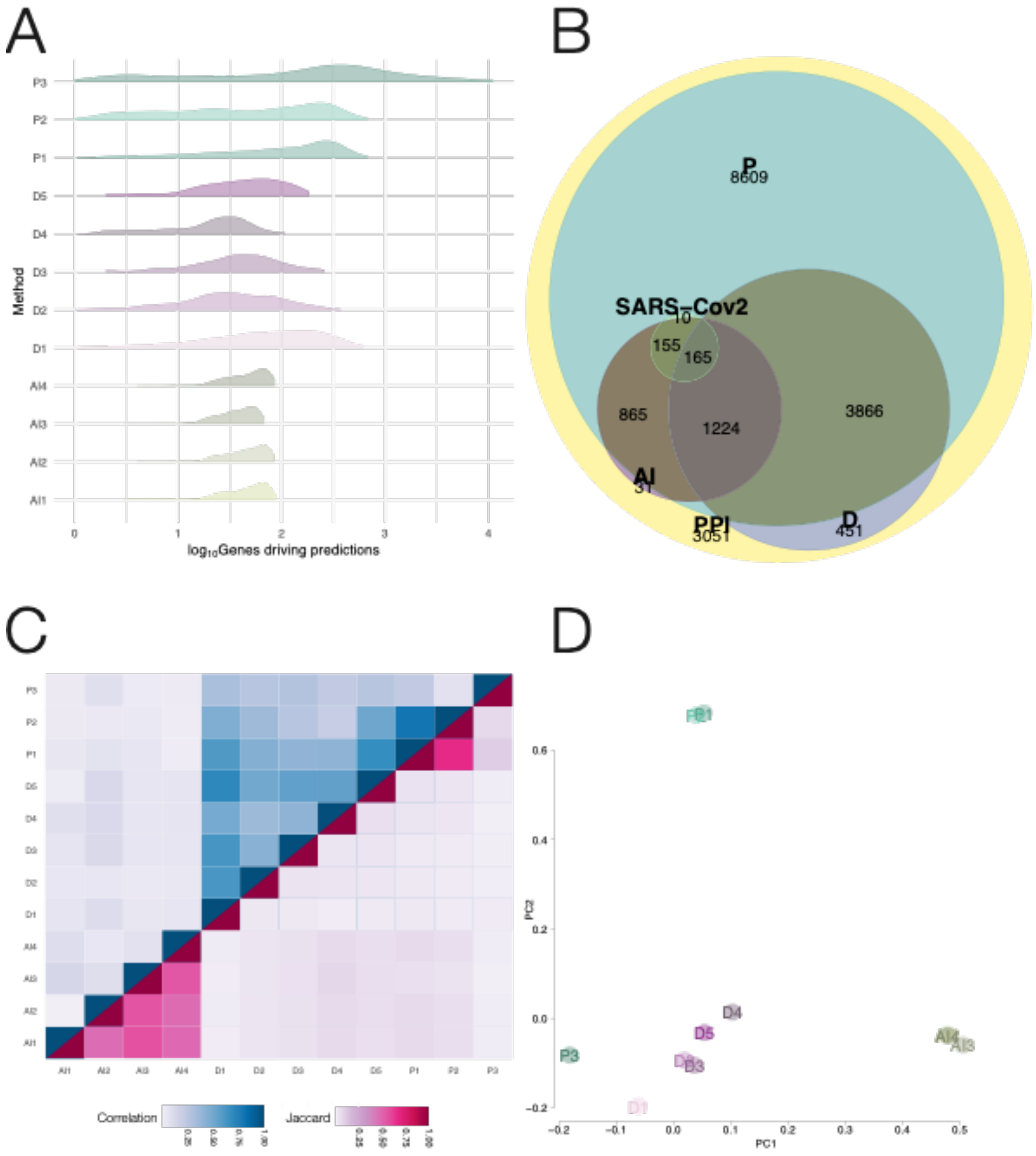

**Figure S4 - Similarities and Differences of the Explanatory Subgraphs.** (A) Distribution of the size of the subnetworks predicted varies according to the method. The AI methods have a smaller variance in the size, while methods based on proximity tend to have higher variances. (B) Gene overlap of the methods involved with subgraphs for each method. Proximity and Diffusion bases methods explore the PPI in a much vast and diverse way than the AI methods (C) Methods inside the same pipeline tend to select similar genes, the similarity of selected genes across methods is different (Jaccard Index), those genes, interestingly, also do not lie in similar neighborhood (similarity), meaning that not only do the genes not overlap across methods, but the vicinity the methods explore are also different. (D) Another measure used to understand methods similarity involved using the PCA of gene drug pairs, showing that AI methods are fairly consistent in what they observe, and similarly, P1 and P2. Diffusion methods have a higher variance in gene-drug pair predictions and have a larger spread of their module; as expected, P3 is far from other proximity measures.

### 1.5.2 Complementarity of Prediction Algorithms (Fig 2C)

To investigate the complementarity among the prediction algorithms, for each drug we measured the network separation $S_{G-d}$ between the explanatory subgraph $G$ and the drug's targets ($d$), and the separation $S_{G-v}$ between $G$ and the 332 SARS-Cov2 viral targets ($v$) capturing the disease module. Each drug has twelve subgraphs, each corresponding to one of the twelve pipelines. A total of 320 drugs, for which all pipelines have predictive subgraph and separation values, are shown in **Figure S*5***. Proximity Pipeline 3 uses differentially expressed genes as input drug data; thus, for proximity P3 we computed the separation between the subgraph and the differentially expressed genes. The figure shows complementarily patterns between methods: the AI pipelines extracts their predictions from subgraphs that overlap with the drug targets ($S_{G-d} < 0$), but are separated from the COVID-19 module ($S_{G-v} > 0$); proximity-based methods show the opposite pattern – for most of the predictive subgraphs the overlap with the COVID-19 module is apparent ($S_{G-v} < 0$); by contrast, diffusion-based predictive subgraphs avoid both the drug targets and the disease module ($S_{G-d} > 0, S_{G-v} > 0$).

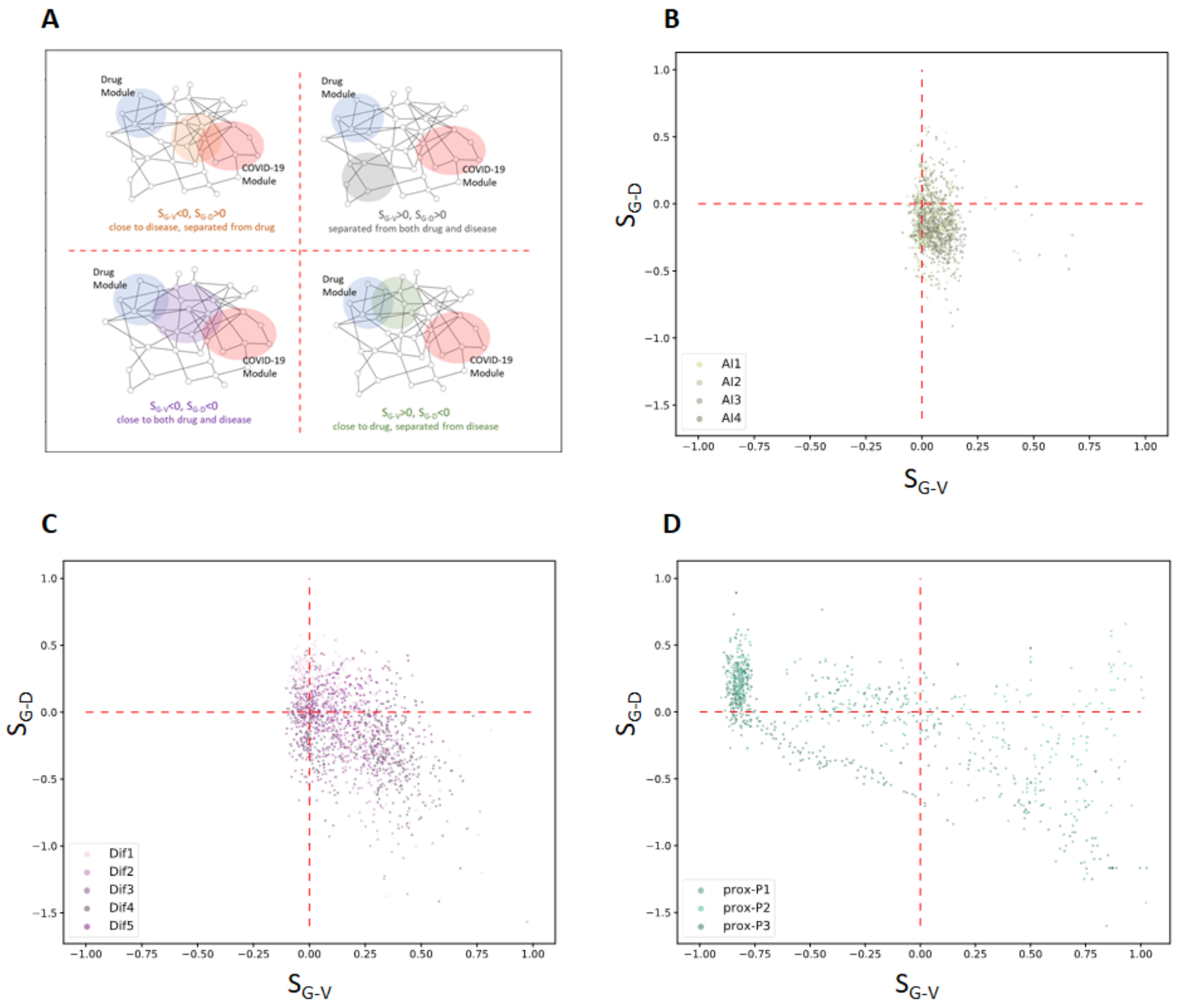

**Figure S5 - The Separation Plot for 320 Drugs.** For each drug, we identified the predictive subgraph for each predictive pipeline. For each subgraph G, we compute the separation between the subgraph G and drug targets as $S_{G\text{-}D}$ and separation between the subgraph and SARS-CoV-2 targets as $S_{G\text{-}V}$. We plot each subgraph as a dot with the two separation values as coordinates to form the plot above. (A) a schematic showing the network pattern represented by each quadrant; (B)-(D): plot for subgraphs in AI, Diffusion, Proximity pipelines, respectively. Each method's subgraphs locate in different regions in the plot, suggesting that they use complimentary regions of the PPI to make predictions.

# 2 Experimental Validation

## 2.1 Cell Cultivation and Viruses Used

VeroE6 cells were obtained from ATCC (Manassas, VA, USA) and maintained in DMEM supplemented with 10% Fetal bovine serum (FBS) at 37°C in a humidified $CO_2$ incubator. The virus strain used was isolated from a traveler returning to Washington State, USA, from Wuhan, China, (USA-WA1/2020) and was obtained from BEI resources (Manassas, VA, USA). The virus stock was passaged twice on Vero cells by challenging the cells at an MOI of less than 0.01 and incubating until cytopathology was seen (typically 3 d after inoculation). A sample of the culture supernatant was sequenced by next gen sequencing (NGS) and was consistent with the original isolate without evidence of other virus or bacterial contaminants. The virus stock was stored at -80°C.

## 2.2 Virus Infection Inhibition Assay

For evaluation of small molecule efficacy against infection with wild type SARS-CoV-2 virus, compounds were first dissolved to 10 mM in DMSO and then diluted into culture medium before addition to cells. The compound stock was added to Vero cells incubated for a minimum of 1 h and then challenged with virus at a MOI of less than 0.2. Dosing ranged from a final concentration of 25 μM down to 0.2 μM in a two-fold dilution series. As a positive control, 5 μM E-64 was used as it was previously reported to inhibit SARS-CoV-2 infection (Hoffman et al. 2020). Negative controls were <0.5% DMSO. After 1-2 day incubation, cells were treated with 10% buffered formalin for at least 6 h, washed in PBS, and virus antigen stained with SARS-CoV-2 specific antibody (Sino Biologicals, MM05) together with Hoechst 33342 dye to stain cell nuclei. Plates were imaged by a Biotek Cytation 1 microscope, and automated image analysis was used to count total number of infected cells and total cell nuclei. CellProfiler software (Broad Institute, MA, USA) was used for image analysis using a customized processing pipeline (available upon request to RAD). Infection efficiency was calculated as the ratio of infected cells to total cell nuclei. Loss of cell nuclei was used to flag treatments suggestive of host cell toxicity. Where possible, EC50 values were calculated using dose-response models fitted by Graphpad Prism software. If a compound was active at the highest dose but an EC50 value could not be calculated due to insufficient activity, the percent inhibition of infection at 25 μM was used to rank potency. Each assay was performed in duplicate in 384 well plates.

## 2.3 Drug-Response Classification

The classification of the drug-response outcomes was done using a drug response curve (DRC) model[45]. We used the R package drc[46] to calculate the DRCs using a log-logistic model with four parameters (hill, IC50, min, and max). Each drug-response was classified in two steps: first inspecting toxicity and later evaluating the drug effect on the inhibition of viral proliferation.

To inspect the cytotoxicity, we first estimated the model parameters using as response variable the nuclei count in the treated cells, normalized by the nuclei count in the controls. We tested the dose-response effect for all drugs using a $\chi^2$ test for goodness of fit and drugs with $p < 0.01$ (Bonferroni correction) were defined as cytotoxicity, with the exception of drugs with toxicity only at the last dose concentration. To evaluate inhibition of viral replication, we used as response for the DRC model the number of infected cells in the treated samples normalized by the controls. For that, a drug was considered to have a dose-response effect by using a $\chi^2$ test for goodness of fit ($p < 0.01$, Bonferroni correction), and the significant drugs were defined as Strong (S) or Weak (W) if the viral reduction was greater than 80% and 50%, respectively. The drugs that did not meet the criteria for S or W were classified as no-effect (N). Finally, we classified drugs as cytotoxic (C) if their toxicity curves were greater than their viral proliferation curves in at least half of the doses tested.

## 2.4 Biological Interpretation of Effective Drugs in E918 Dataset

We observed 77 drugs that showed strong (S) or weak effects (W) in the high-throughput screening. There was no drug category (ATC Classification) that was enriched among the S, W, or S&W drugs (hypergeometric test FDR-BH padj > 0.05). To search for common patterns that could explain their bioactivity, we performed hierarchical clustering on the drug target profiles, failing to find binding patterns shared by all drugs (*Figure S6*). Only four small groups of drugs are observed, documenting various degrees of shared targets (*Figure S6*), three of which contain drugs from multiple categories, and one group consists of 7 nervous system-related drugs with similar target profiles. We also performed pathway enrichment analysis to identify biological processes shared across the targets of drugs with strong or weak effects. Among the 77 S&W drugs, 42 are located in three groups associated with common pathways, and 20 of these drugs are of diverse indications linked to transport and metabolism of different substrates. Eighteen are associated with pathways related to membrane receptors, most of them indicated for nervous system disorders, targeting G protein-coupled receptors such as ADRA1A, HTR2A, and HRH1 (*Figure S7*). Taken together, neither the pathway nor the target analysis reveals patterns that could explain the efficacy of the 77 S&W drugs.

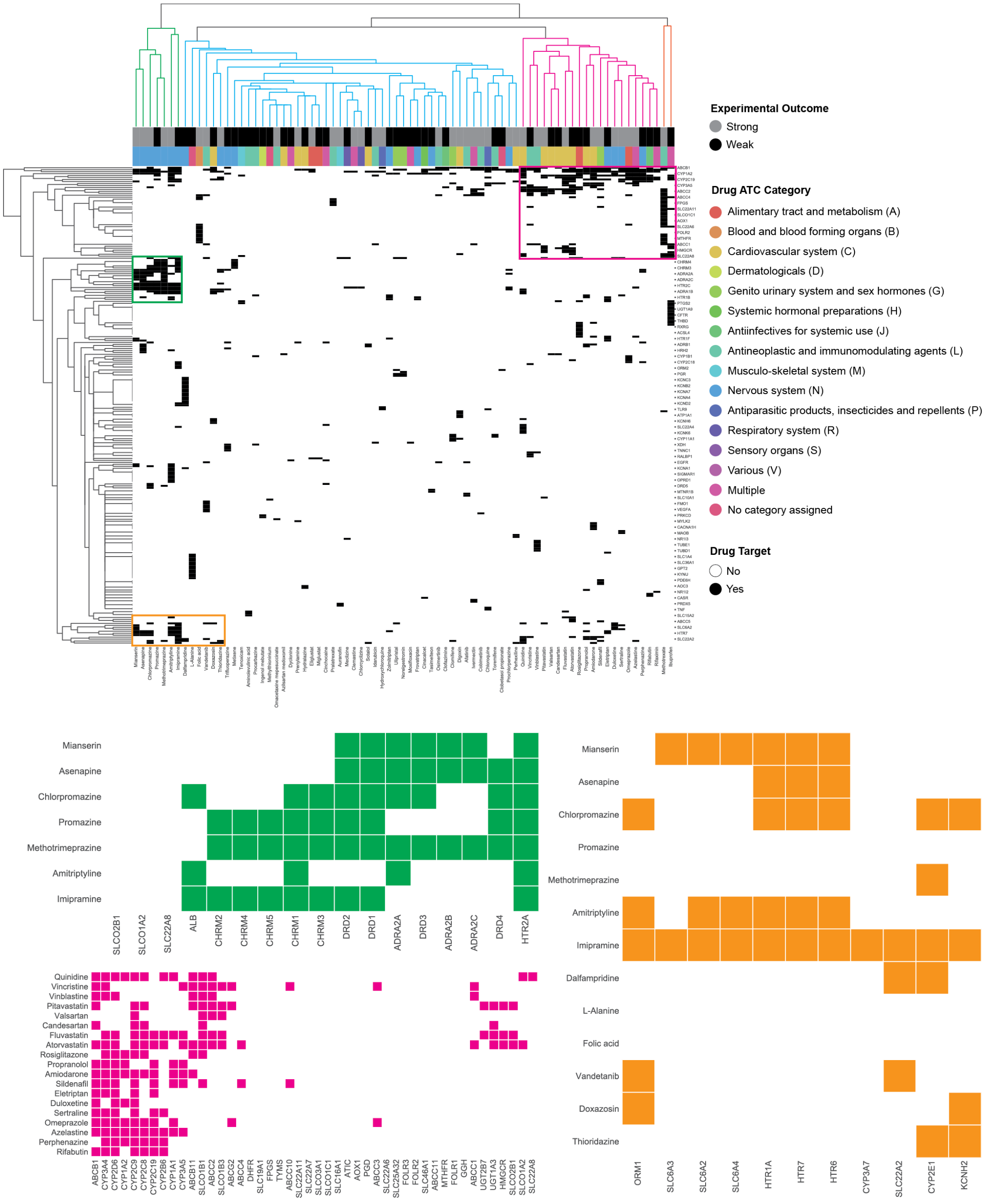

**Figure S6 – Hierarchical Clustering Highlights Groups of Drugs with Similar Target Profiles**. Heatmap showing 77 S&W drugs from the E918 dataset and their respective targets (colored cells). Clustering performed using Euclidean distance and single linkage.

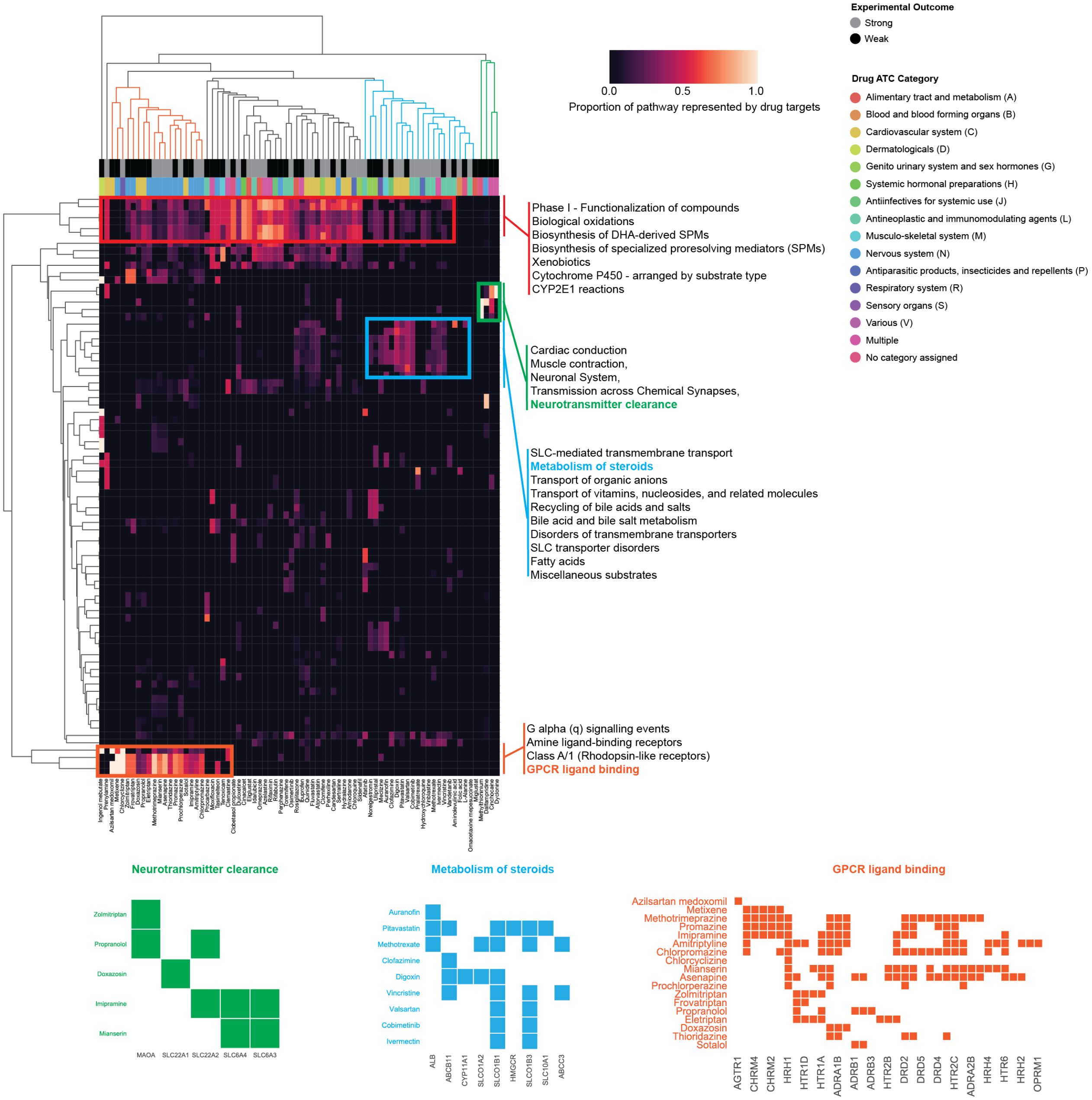

**Figure S7 – Pathway Enrichment.** Heatmap showing successful (S&W) drugs in the E918 dataset and their respective Reactome pathways in which their targets are enriched. Hierarchical clustering (Euclidean, single linkage) highlights different groups of drugs with similar pathway profiles. We highlight the pathways for three drug clusters, emphasizing the proteins targeted in one exemplary pathway for each cluster

# 3 Statistical Validation

## 3.1 Performance Evaluation using ROC Curves, Precision, and Recall

We examined whether positive drugs (e.g., strong-effect drugs) were ranked high by measuring the predictive power of each pipeline in terms of area under the ROC (Receiver Operating Characteristics) curve, precision, and recall. First, we calculated ROC (Receiver Operating Characteristics) curves and AUC (area under the curve) scores for model selection and performance analysis. The AUC score measures the separation between positive examples (e.g., drugs with strong or weak responses) and negative examples (e.g., drugs showing no-effect in experimental screening). For the ranked lists of drugs, we applied different thresholds to compute false-positive and true-positive rates to plot the ROC curves. Scores of AUC range between 0 and 1, where 1 corresponds to perfect performance and 0.5 indicates the performance of a random classifier. We used the Python package Scikit-learn[47] for computing the AUC scores and plotting the ROC curves.

The AUC metric operates on the whole ranked list of drugs, and thus it does not directly reflect the ability of the method to prioritize most promising drug candidates at the top of the list. To address this issue and account for unbalanced ground-truth information where negative examples vastly outnumber positives, we also considered hit-rate based metrics to evaluate the quality of top-K drugs in each ranked list. Here, we evaluated performance at a given cut-off rank K, considering only the topmost predictions by the pipeline. In particular, we calculated the fraction of top-K ranked drugs that were positive outcomes (precision at K) and the fraction of all positive outcomes that were among the top-K ranked drugs (recall at K).

We considered three types of ground-truth information to evaluate prediction performance:

1) The outcome of the experimental screening of 918 compounds (E918 dataset, Table S10). We identified 806 no effect drugs, 40 with weak effect, and 37 with strong effect.
2) The outcome of the experimental screening of additional 74 compounds tested with a wider range of doses (0.625 – 20μM, 0.2 MOI) (E74 dataset, Table S11) (Figure S8). The E74 dataset represents a subset of 81 compounds by a medical doctor among the top 10% of all drug predictions that were available for purchase. We identified 39 no effect drugs, 10 with weak effect, and 11 with strong effect.

3) 67 drugs that, as of April 2020, were in ongoing trials for COVID-19, obtained from the ClinicalTrials.gov website** (CT415 dataset, Table S12). ClinicalTrials.gov organizes COVID-19 specific collection of all trials. Trial records consist of information on inclusion and exclusion criteria, details on drugs being tested, the scientific team behind the study, and funding agencies. We extract drug names from clinical trials' treatment information and match their names with records on the DrugBank database[20].

4) We also collected clinical trials data at the experimental readout time 6/15/2020 (C615 dataset) (Table S13).

Note that some methods do not provide prediction for every drug in the full dataset. While that would make a fair comparison of the methods challenging, we note that ground-truth information described above is available for drugs predicted by all pipelines (except for P3, hence it is harder to compare this pipeline with the other 11). Finally, we note that we adopted a conservative approach by evaluating predictive performance using the rankings across all 6,340 drugs, not only 918 experimentally screened drugs. For example, it is possible to conceive that a particular topmost prediction in a pipeline represents a positive drug, however, that is impossible to know if the predicted drug was not included in experimental screening. Because of that, the reported precision and recall values represent conservative estimates of prediction performance, i.e., the values are lower than what one could obtain if the analysis was limited to only experimentally screened drugs. To determine the significance of predictive power, we calculated the expected number of positive drugs among top-K drugs for each pipeline and compared the expected values with the observed precision and recall values. To this end, we calculated the expected number of positive drugs by taking into account (a) the number of drugs for which ground-truth information is available, and (b) the number of drugs for which a pipeline makes predictions. We used an exact one-tailed binomial test ($p$-value $< 0.05$) to test whether a top-K list returned by a pipeline is biased towards containing more positive drugs than what we would expect on average by pure chance had the ranking be a random one.

** Clinical Trial Covid-19 selection: https://clinicaltrials.gov/ct2/results?cond=COVID-19

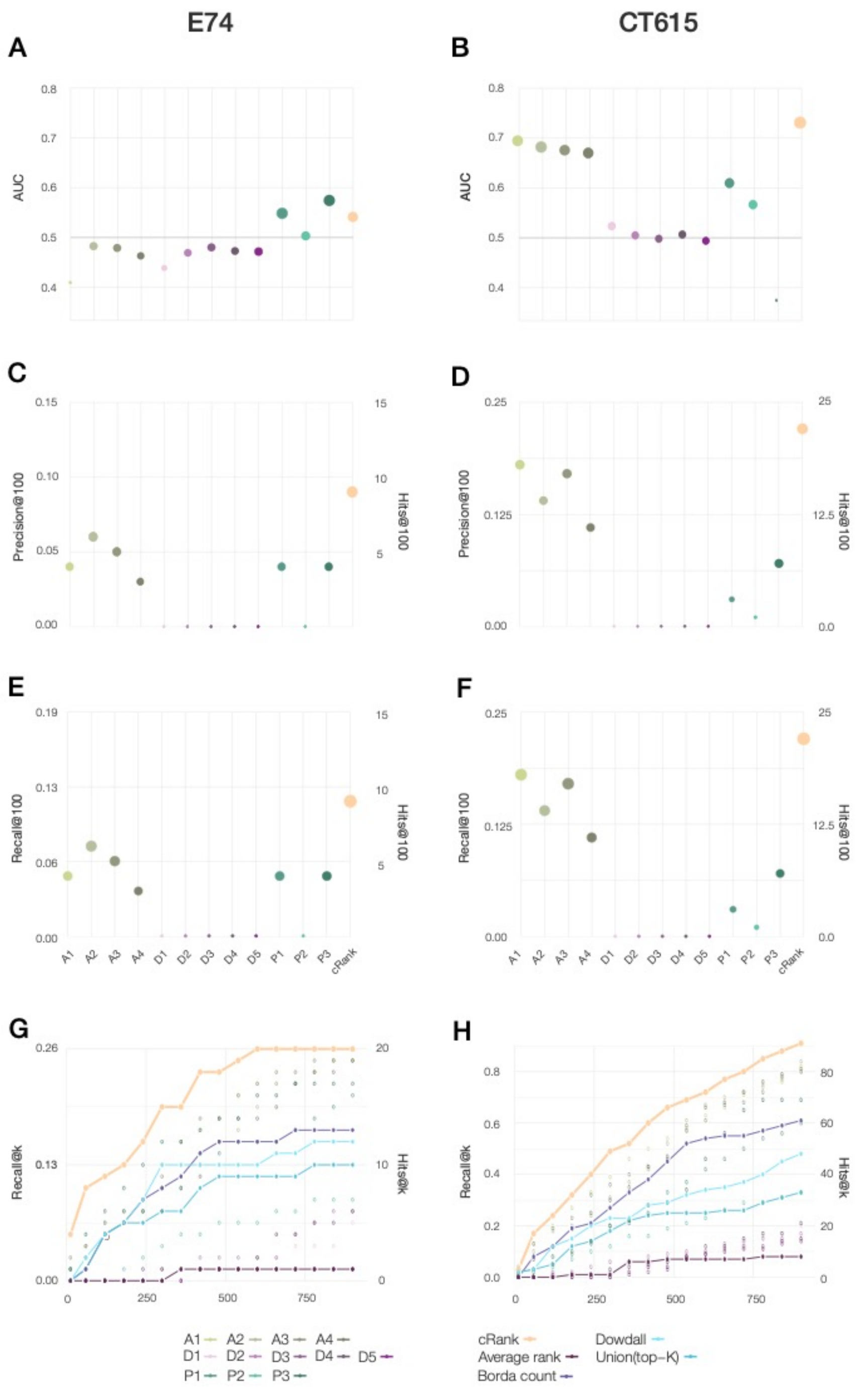

**Figure S8 - Performance of the Different Predictive Pipelines.** (**A,B**) AUC (Area under the Curve), (**C,D**) precision at 100, and (**E,F**) recall, for twelve pipelines tested for drug repurposing, using as a gold standard the S&W drugs in E74 (left panel, experimentally validated dataset from expert curation and drug selection) and CT615 (right panel, drugs in clinical trials until July 15th 2020). (**G,H**) The top precision and recall for the different rank aggregation methods (connected points), compared to the individual pipelines (empty symbols) documenting the strong predictive performance of CRank. CT06 presents, in most cases, higher hit rates, precision and recalls when compared to E74.

Table S1 – Performance evaluation using the E918 dataset as ground-truth (Fig 4C,E)

| | Precision@100 | Recall@100 |
|---|---|---|
| **CRank** | 9% (0.750%, $p=7.01 \times 10^{-8}$) | 11.7% (0.9%, $p=2.61 \times 10^{-8}$) |
| **P1** | 7% (0.467%, $p=3.77 \times 10^{-7}$) | 9.1% (0.6%, $p=2.52 \times 10^{-7}$) |
| **P2** | 0% (0.458%, p=1.00) | 0% (0.5%, p=1.00) |
| **P3** | 5% (2.453%, p=0.090) | 6.5% (2.3%, p= 0.031) |
| **D1** | 2% (0.062%, p=0.001) | 2.6% (0.1%, $p=9.32 \times 10^{-4}$) |
| **D2** | 2% (0.075%, p=0.002) | 2.6% (0.1%, $p=2.29 \times 10^{-3}$) |
| **D3** | 1% (0.062%, p=0.061) | 1.3% (0.1%, p=0.061) |
| **D4** | 1% (0.037%, p=0.037) | 1.3% (0.1%, p=0.037) |
| **D5** | 1% (0.062%, p=0.061) | 1.3% (0.1%, p=0.061) |
| **AI1** | 4% (2.779%, p=0.302) | 5.2% (3.3%, p=0.024) |
| **AI2** | 9% (2.922%, p=0.002) | 11.7% (3.7%, $p=2.14 \times 10^{-3}$) |
| **AI3** | 6% (2.635%, p=0.047) | 7.8% (3.2%, p=0.037) |
| **AI4** | 6% (2.922%, p=0.049) | 7.8% (3.5%, p=0.054) |

The values in the brackets represent expected values, i.e. expected Recall@100 or expected Precision@100, followed by p-values

Table S2 – Performance evaluation using the CT415 dataset (Fig 4D,F)

| | Precision@100 | Recall@100 |
|---|---|---|
| **CRank** | 12% (0.360%, $p=3.06 \times 10^{-15}$) | 32.4% (0.5%, $p=2.17 \times 10^{-19}$) |
| **P1** | 1% (0.224%, p=0.201) | 2.7% (0.6%, p=0.202) |
| **P2** | 0% (0.220%, p=1.00) | 0.0% (0.6%, p=1.00) |
| **P3** | 2% (1.790%, p=0.538) | 5.4% (2.6%, p=0.245) |
| **D1** | 0% (0.030%, p=1.0) | 0% (0.1%, p=1.00) |
| **D2** | 0% (0.036%, p=1.0) | 0% (0.1%, p=1.00) |
| **D3** | 0% (0.030%, p=1.0) | 0% (0.1%, p=1.00) |
| **D4** | 0% (0.018%, p=1.0) | 0% (0.1%, p=1.00) |
| **D5** | 0% (0.302%, p=1.0) | 0% (0.1%, p=1.00) |
| **AI1** | 12% (1.335%, $p=7.39 \times 10^{-9}$) | 32.4% (2.3%, $p=2.41 \times 10^{-11}$) |
| **AI2** | 10% (1.404%, $p=1.29 \times 10^{-6}$) | 27.0% (2.3%, $p=8.25 \times 10^{-9}$) |
| **AI3** | 11% (1.267%, $p=4.52x \times 10^{-8}$) | 29.7% (2.3%, $p=4.74 \times 10^{-10}$) |
| **AI4** | 9% (1.404%, $p=1.06 \times 10^{-5}$) | 24.3% (2.3%, $p=1.26 \times 10^{-7}$) |

The values in the brackets represent expected values, i.e. expected Recall@100 or expected Precision@100, followed by p-values

# 4 Rank Aggregation Algorithms (RAAs)

Rank aggregation is concerned with how to combine several independently constructed rankings into one final ranking that represents a consensus ranking, i.e., a collective opinion of prediction methods that is representative of all rankings returned by the methods[48]. The classical consideration for specifying the final ranking is to maximize the number of pairwise agreements between the final ranking and each input ranking. Unfortunately, this objective, known as the Kemeny consensus, is NP-hard to compute[48,49], which has motivated the development of methods that either use heuristics or approximate the Kemeny optimal ranking[48,50–52].

## 4.1 Average Rank Method

The Average Rank method follows the most straightforward way to integrate multiple rankings. For each drug, it calculates a simple rank average over 12 rankings returned by the pipelines to obtain the overall ranking. While the Average Rank method is a popular ad-hoc rank aggregation strategy, many studies[53–55], including ours, found that studying the average ranks can be a poor aggregation approach. Next, we briefly overview methods that realize more sophisticated approaches to obtain the overall ranking.

## 4.2 Borda Method

The Borda method[56] is one of most commonly used rank aggregation methods. Briefly, the method proceeds as follows. Given are $k$ rankings exist, $R_1, R_2, \ldots, R_k$. For each drug $a \in R_i$, $a$ is assigned a score $B_i(a)$ equal to the number of drugs that $a$ outranks in ranking $R_i$. The Borda count $B(a)$ of drug $a$ is then calculated as $\sum_{i=1}^{k} B_i(a)$. Finally, drugs are sorted in the descending order based on their Borda counts to create a consensus ranking. Theoretically, Borda method offers a guarantee on approximating Kemeny consensus. In particular, Borda method is a 5-approximation algorithm of the Kemeny optimal ranking[51].

## 4.3 Dowdall Method

The Dowdall method[57] is a modified form of the Borda method that has been widely used in political elections in many countries. Intuitively, individual pipelines make predictions for drugs, which are interpreted as preferences of the pipeline. For a pipeline, its 1st choice gets a score of 1, its 2nd choice get 1/2, its 3rd choice gets 1/3, and so on. Drug with the largest total score across pipelines wins. Formally, let be given $k$ rankings, $R_1, R_2, \dots, R_k$. For each drug $a \in R_i$, $a$ is first assigned a score $D_i(a)$ equal to the reciprocal of drug's rank in ranking $R_i$. The total score $D(a)$ is then calculated as $\sum_{i=1}^{k} D_i(a)$. Candidates are sorted in descending order based on their total score to create a consensus ranking.

## 4.4 CRank

The CRank algorithm[58] starts with ranked lists of drugs, $R_r$, each one arising from a different pipeline, $r$. Each ranked list is partitioned into equally sized groups, called bags. Each bag $i$ in ranked list $R_r$ has attached importance weight $K_r^i$ whose initial values are all equal. CRank uses a two-stage iterative procedure to aggregate the individual rankings by taking into account uncertainty that is present across ranked lists. After initializing the aggregate ranking $R$ as a weighted average of ranked lists $R_r$, CRank alternates between the following two stages until no changes were observed in the aggregated ranking $R$. (1) First, it uses the current aggregated ranking $R$ to update the importance weights $K_r^i$ for each ranked list. For that purpose, the top-ranked drugs in $R$ serve as a temporary gold standard. Given bag $i$ and ranked list $R_r$, CRank updates importance weight $K_r^i$ based on how many drugs from the temporary gold standard appear in bag $i$ using Bayes factors[59,60]. (2) Second, the ranked lists are re-aggregated based on the importance weights calculated in the previous stage. The updated importance weights are used to revise $R$ in which the new rank $R(a)$ of drug $a$ is expressed as: $R(a) = \sum_r K_r^{i_r(a)} R_r(a)$, where $K_r^{i_r(a)}$ indicates the importance weight of bag $i_r(a)$ of drug $a$ for ranking $r$, and $R_r(a)$ is the rank of $a$ according to $r$. By using an iterative approach, CRank allows for the importance of a ranked list returned by an individual pipeline not to be predetermined, i.e., a-priori fixed, and to vary across drugs. The final output is a global ranked list $R$ of drugs that represents the collective opinion of all drug repurposing prediction algorithms. In all experiments, we set the number of bags to 1,000, the size of the temporary gold standard to 0.5% of the total number of drugs in $R$, and the maximum number of iterations to 50. In all cases, the algorithm converged, in fewer than 20 iterations[59,60]. The pipelines' ranked lists and CRank's aggregation are provided in Table S14. The Python source code implementation of CRank is available at https://github.com/mims-harvard/crank (raa.py).

### 4.5 Comparison of RAAs

What explains CRank's outstanding performance across all datasets? Each RAA aims to approximate the optimal Kemeny consensus, which offers the best agreement with all 12 prediction pipelines. As this consensus remains unknown (NP-hard), we cannot assess how well the different RAA methods approximate it. We do, however, have a ground-truth ranking, offered by the experimental and clinical datasets (E918 and CT415). We assigned rank 1 to the strong drugs, rank 2 to the weak drugs, and rank 3 to the no-effect drugs, allowing us to measure the Kemeny score for each aggregated list, representing the fraction of pairwise disagreements between the respective ranked list and the experimental outcomes. For $\mathrm{K} = 100$, the Kemeny score of the Average Rank method is infinite for E918, as there are no positive drugs among the top 100. In contrast, for the Borda count, we obtain a Kemeny score of KS = 0.7131, indicating that 71% of all drug pairs in the ranked list of Borda method disagrees with the ground-truth ranking in the E918 dataset. Note that the theoretical expectation for a purely random ranking is KS = 0.5, meaning that 50% of all drug pairs in the random reference are flipped, i.e., while with KS = 0.4545 Dowdall does better than random, we observe a much lower KS = 0.2679 for CRank. We measured the Kemeny score for multiple values, for both datasets (E918 and CT415), finding that for K<250 (top drugs), CRank offers the best agreement with the outcomes.

# 5 Supplementary Tables

Table S3 – Protein-Protein Human Interactome. 332,749 pairwise binding interactions between 18,508 human proteins.

Table S4 - SARS-COV2-Human Interactome. Protein-protein interactions between 29 SARS-COV2 proteins and 332 human proteins detected by affinity purification followed by mass spectrometry (dataset retrieved from Gordon et al (2020)).

Table S5 – List of drugs and their respective targets retrieved from the DrugBank database.

Table S6 – List of 17,222 differentially expressed genes identified by exposure of 793 drugs in different cell lines. Data obtained from the DrugBank database.

Table S7 - Network Overlap Between 299 Diseases and SARS-COV2 Targets. The $S_{vb}$ measure captures the network-based overlap between SARS-COV2 targets $v$ and the gene pool associated with disease $b$.

Table S8 - Embedding vectors. Representations of diseases as learned by the GNN model. Each row in the file contains the embedding vector for a particular disease.

Table S9 - Embedding vectors. Representations of drugs as learned by the GNN model. Each row in the file contains the embedding vector for a particular drug.

Table S10 – The E918 dataset. List of 918 drugs screened for their efficacy in inhibiting SARS-CoV-2 in VeroE6 cells and their experimental outcome.

Table S11 - The E74 dataset. Experimental outcomes of 74 compounds selected by a medical doctor among the top 10% of all drug predictions that were available for purchase.

Table S12 - Drugs Under Evaluation in Clinical Trials for Treating COVID-19 (as of April 2020) (C415 dataset).

Table S13 - Drugs Under Evaluation in Clinical Trials for Treating COVID-19 (as of June 2020) (C615 dataset).

Table S14 – Drug Rankings. Ranking of each drug obtained by the 12 pipelines and their aggregation with CRank.